\documentclass[fleqn,usenatbib]{mnras}
\usepackage{newtxtext,newtxmath}
\usepackage[T1]{fontenc}
\usepackage{lipsum}
\DeclareRobustCommand{\VAN}[3]{#2}
\let\VANthebibliography\thebibliography
\def\thebibliography{\DeclareRobustCommand{\VAN}[3]{##3}\VANthebibliography}

\usepackage{subfigure}
\usepackage{gensymb}
\usepackage{multicol}
\usepackage{dblfloatfix} 
\usepackage{multirow}
\usepackage{booktabs}
\usepackage{graphicx}	% Including figure files
\usepackage{amsmath}	% Advanced maths commands
\newcommand{\kms}{km s$^{-1}$}
\title[HexPak]{SDSS-IV MaNGA: Calibration of astrophysical line-widths in the H$\rm\upalpha$ region using HexPak observations}

\author[S. Chattopadhyay et al.]{
Sabyasachi Chattopadhyay,$^{1}$\rm\thanks{E-mail: sabyasachi@saao.ac.za} Matthew A. Bershady,$^{1,2,3}$ David R. Law,$^4$ 
Kyle Westfall,$^5$ Shravan Shetty,$^{2,6}$ 
\newauthor Camilo Machuca,$^2$ Michele Cappellari,$^7$ Kate H. R. Rubin,$^{8,9}$ Kevin Bundy,$^5$ Samantha Penny$^{10}$  
\\
$^{1}$South African Astronomical Observatory, 1 Observatory Rd, Observatory, Cape Town, 7925, South Africa\\
$^{2}$University of Wisconsin, Department of Astronomy, 475 North Charter Street, Madison, WI 53706, USA\\
$^{3}$Department of Astronomy, University of Cape Town, Private Bag X3, Rondebosch 7701, South Africa\\
$^4$Space Telescope Science Institute, 3700 San Martin Drive, Baltimore, MD 21218, USA\\
$^5$University of California Observatories, University of California,
Santa Cruz, 1156 High St., Santa Cruz, CA 95064, USA\\
$^6$Kavli Institute for Astronomy and Astrophysics, Peking University, Beijing 100871, China \\
$^7$Sub-department of Astrophysics, Department of Physics, University of Oxford, Denys Wilkinson Building, Keble Road, Oxford OX1 3RH, UK\\
$^8$Department of Astronomy, San Diego State University, San Diego, CA 92182 USA \\
$^9$Center for Astrophysics and Space Sciences, University of California, San Diego, La Jolla, CA 92093, USA \\
$^{10}$ Institute of Cosmology \& Gravitation, University of Portsmouth, Dennis Sciama Building, Portsmouth, PO1 3FX, UK
}

\date{Accepted XXX. Received YYY; in original form ZZZ}

\pubyear{2023}

\begin{document}
\label{firstpage}
\pagerange{\pageref{firstpage}--\pageref{lastpage}}
\maketitle

\begin{abstract}
We have re-observed $\rm\sim$40 low-inclination, star-forming galaxies from the MaNGA survey ($\upsigma\sim65$~\kms) at $\sim$6.5 times higher spectral resolution ($\upsigma\sim10$~\kms) using the HexPak integral field unit on the WIYN 3.5m telescope. The aim of these observations is to calibrate MaNGA's instrumental resolution and to characterize turbulence in the warm interstellar medium and ionized galactic outflows. Here we report the results for the H$\rm\upalpha$ region observations as they pertain to the calibration of MaNGA's spectral resolution. Remarkably, we find that the previously-reported MaNGA line-spread-function (LSF) Gaussian width is systematically underestimated by only 1\%. The LSF increase modestly reduces the characteristic dispersion of HII regions-dominated spectra sampled at 1-2 kpc spatial scales from 23 to 20 km s$^{-1}$ in our sample, or a 25\% decrease in the random-motion kinetic energy. This commensurately lowers the dispersion zeropoint in the relation between line-width and star-formation rate surface-density in galaxies sampled on the same spatial scale. This modest zero-point shift does not appear to alter the power-law slope in the relation between line-width and star-formation rate surface-density. We also show that adopting a scheme whereby corrected line-widths are computed as the square root of the median of the difference in the squared measured line width and the squared LSF Gaussian
% as a median statistic of squared differences 
avoids biases and allows for lower SNR data to be used reliably.

\end{abstract}


%%%%%%%%%%%%%%%%%%%%%%%%%%%%%%%%%%%%%%%%%%%%%%%%%%

%%%%%%%%%%%%%%%%% BODY OF PAPER %%%%%%%%%%%%%%%%%%

\section{Introduction}
\label{sec:intro}

Spectroscopic determination of gas and stellar motions -- kinematics -- is a fundamental diagnostic of physical conditions in the interstellar medium (ISM) and stellar dynamics, respectively. In SDSS-IV \citep{Blanton_2017}, the BOSS spectrographs \citep{Smee_2013} on the Sloan 2.5m Telescope \citep{Gunn_2006} were retrofitted with a suite of positionable, multi-object integral field units \citep[IFUs;][]{Drory_2015} to conduct a large spectroscopic survey of nearby galaxies: Mapping Nearby Galaxies at APO survey \cite[MaNGA;][]{Bundy_2015, Law_2015, Yan_2016a, Yan_2016b, Wake_2017}. MaNGA's unsurpassed spectral coverage (360-1000~nm) and sample size (more than 10,000 galaxies) provide an unprecedented data-set for analysis of the spatial distribution and properties of stellar populations \cite[e.g.,][]{Neumann_2022,Sanchez_2022} as well as the physical conditions of the ionized gas \cite[e.g.,][]{Mingozzi_2020,Schaefer_2020,Belfiore_2016} in galaxies over several decades in mass. However, the BOSS spectrographs with the MaNGA fibres are limited to a spectral resolution of $\sim$2000 equivalent to $\rm\sim$70 km s$^{-1}$ ($\rm\upsigma$), several times larger than the intrinsic line-widths of gas and stars in normal, star-forming spiral disks and dwarf galaxies \citep[$\rm\upsigma\sim$ 10 to 20~km~s$^{-1}$, e.g.,][]{Terlevich_1981, Melnick_88, Bender_92, Andersen_2006, Epinat_08, Martinsson_2013, Penny_15}. As a consequence of this instrumental limitation there has been relatively little exploration of disc stellar dynamics with MaNGA data, and in fact those studies that do have turned to proxies such as asymmetric drift \cite[][]{Shetty_2020, Meng_2021}.

To circumvent these instrumental limitations on the MaNGA data application the MaNGA survey team have undertaken a major effort to characterize the BOSS spectrograph instrumental resolution \cite[the line-spread function, or LSF;][]{Law_2016, Law_2021} as part of the deliverable science products provided by MaNGA's Data Analysis Pipeline \citep{Westfall_2019,Belfiore_2019}. The quality of this characterization is of such high fidelity that current estimates of the LSF provide corrections approaching what is required to achieve reliable line-widths at the $\rm\upsigma\sim$20 to 30~km~s$^{-1}$ level for individual emission-lines. The effort promptly yielded two significant scientific results on the nature of the ionized gas in galaxies concerning  the correlation of (i) line-ratios and line-widths \citep{Law_2021b} that separates HII-like from diffused ionized gas; and (ii) line-widths to star-formation rate \citep{Law_2022} that provides a definitive local calibration of star-formation feedback-driven turbulence.

\begin{figure*}
\includegraphics[width=\textwidth]{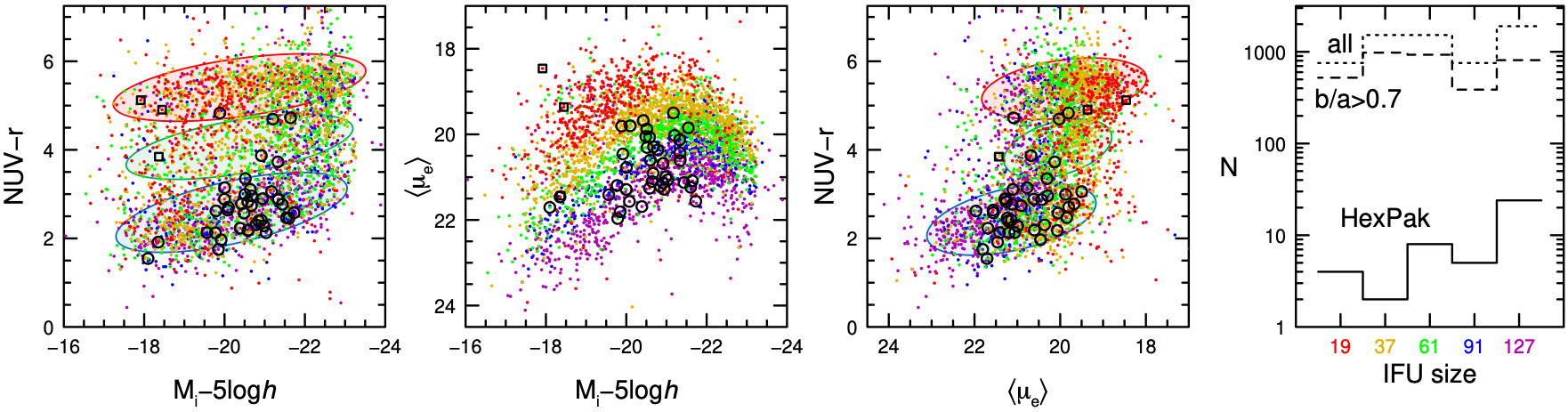} 
\caption{Distribution of HexPak targets (black open symbols, see text) in NUV-r rest-frame color, $i$-band absolute magnitude (\textit{AB} mag), and mean $i$-band surface-brightness (\textit{AB} mag arcsec$^{-2}$) within the elliptical half-light radius (see Table~\ref{tab:sample}) overlaid on the parent MaNGA MPL-8 sample for 3628 galaxies with b/a$>$0.7. The parent sample is coloured by IFU size (fibre count), coded in the right-hand histogram. Red, blue and green ellipses in the first and third panels mark the approximate respective locations of the red galaxy sequence, the star-forming main sequence (or blue cloud), and a transition region (i.e., the green valley).}
\label{fig:hex_sample}
\end{figure*} 

However, the MaNGA LSF calibration hinges on direct comparison with a reference set of a relatively small number of observations at higher spectral resolution. 
 The largest high-resolution comparison is the DiskMass Survey \citep{Bershady_2010a,Bershady_2010b, Westfall_2011, Westfall_2014, Martinsson_2013}, which provides line-widths for H$\rm\upalpha$ nebular emission as well for integrated star-light. However, the overlap of DiskMass and MaNGA is only seven galaxies.
The DiskMass Survey H$\rm\upalpha$ spectra have $R\equiv\lambda/\delta\lambda\sim 10000$ ($\rm\upsigma\sim9.9$ km s$^{-1}$). 
% The DMS Ha sigmas, as well as the earlier DensePak data were corrected for instrumental broadening, but all sigma measurements are post-pixelized values. 
Given the importance of understanding the fidelity of these corrections for a wide range of science applications, we have undertaken a new calibration of the MaNGA LSF by re-observing 43 galaxies from the MaNGA sample in H$\rm\upalpha$ at even higher spectral resolution with a different spectrograph, IFU, and telescope. These new, high-resolution data were taken not only for  purposes of calibrating the MaNGA LSF, but also to determine the impact of the MaNGA LSF on detecting galactic winds. The latter will be reported in a separate publication (Chattopadhyay et al., in preparation).

This paper is organized as follows. Section~\ref{sec:data} presents the sample selection and describes the new data acquired with the HexPak IFU at the WIYN 3.5m telescope along with the basic image processing and spectral extraction. Section~\ref{sec:measurements} details measurements of the WIYN Bench Spectrograph LSF using two different tracers, and the spatial registration of MaNGA and HexPak IFUs. In Section~\ref{sec:analysis} we compare the HexPak and MaNGA H$\rm\upalpha$ line-widths, corrected for instrumental broadening. We use the systematic differences between these line-widths to derive a correction to the MaNGA  LSF estimate. Section~\ref{sec:results} provides illustrations of the scientific impact of this correction to the MaNGA LSF. In Section~\ref{sec:conclusions}  we summarize our conclusions from these sections in the context of the HII-like line-width distribution and the correlation between H$\rm\upalpha$ line-width and star-formation surface-density. Throughout, we follow the nomenclature from \cite{Law_2021} where $\rm\upsigma$ refers to the observed or corrected Gaussian line-width of an astrophysical source, while $\rm\upomega$ refers to the Gaussian profile characterization of the spectrograph LSF as measured from the widths of monochromatic arc or sky lines. Line wavelengths are specified in air.

\section{Data}
\label{sec:data}

\subsection{Sample selection from MaNGA}
\label{sample:seletion}

We targeted low-inclination galaxies with axis ratios b/a$>$0.7 from galaxies observed by and included in the MaNGA Project Launch-8 (MPL-8), released internally in November 2018. This parent sample consists of datacubes for 6430 galaxies, post-dates the public Data Release (DR)-15 \citep{DR15} and predates DR-17 \citep{DR17}, but is fully contained in the latter. Low-inclination galaxies were selected to minimize line-of-sight effects and to be sensitive to winds. We also inspected all relevant datacubes to ensure that they had reasonable kinematic signal in ionized gas (moderate to high star-formation) and integrated star-light (continuum surface-brightness above 22 mag arcsec$^{-2}$ within the half-light radius in the $i$ band). 

Due to instrument constraints on spectral coverage at high resolution, source redshifts between $0.018<z<0.028$ were required to ensure [S~II]$\rm\lambda\lambda671.7,673.0$~nm was captured in the echelle, order 8 setup for the H$\rm\upalpha$ region (this set the upper redshift limit), while [O~III]$\rm\lambda$500.7~nm was captured in the VPHg setup for the Mgb region (this set the lower redshift limit).

All considerations being equal, we preferentially observed targets with HexPak that had been observed in the larger MaNGA IFUs. This was for the practical reason that the HexPak IFU has a larger footprint that the MaNGA IFUs, while the inner 4 arcsec (radius) of the MaNGA data has significant beam-smearing \citep{Law_2021}; the larger MaNGA IFUs therefore provide the greatest spatial overlap where kinematics can be well compared. We note that this preferentially samples lower surface-brightness galaxies \citep{Wake_2017}, as seen in Figure~\ref{fig:hex_sample} here. However, we do sample galaxies over a wide range of total star-formation from 0.05 to 3 M$_\odot$ yr$^{-1}$ (H$_0 = 70$ km s$^{-1}$ Mpc$^{-1}$).

A target table with salient data is given in Appendix~\ref{app:targets}, Table~\ref{tab:sample}. Figure~\ref{fig:hex_sample} shows the distribution of HexPak targets in rest-frame color, luminosity and surface-brightness within the larger MaNGA low-inclination sample (MPL-8; b/a$<$0.7). The distribution shows our preference for blue star-forming galaxies observed with larger MaNGA IFU sizes, and within these constraints, with higher surface-brightness.

Finally, three low-mass AGN hosts, selected from \cite{Penny_2018}, were observed in the first semester of the program: 1-230177, 1-379255, 1-38618. These are marked with open squares in Figure~\ref{fig:hex_sample}. None of these targets exhibited strong H$\rm\upalpha$ emission in the HexPak data, and had very little radial extent. These galaxies have been excluded from the remaining analysis.

\subsection{New Observations: HexPak}
\label{sec:hex}

The program was awarded 20 nights on the WIYN 3.5m telescope\footnote{The WIYN Observatory is a joint facility of the University of Wisconsin–Madison, Indiana University, NSF’s NOIRLab, the Pennsylvania State University, Purdue University, and the University of California, Irvine.} using the HexPak IFU \citep{Wood_2012} and the upgraded Bench Spectrograph \citep{Bershady_2008, Knezek_2010} over three observing semester starting in the second half of 2018. HexPak is roughly a 41 by 36 arcsec hexagon of 84 $\rm\times$ 2.81 arcsec diameter fibres with a 6 arcsec diameter core of 18 $\rm\times$ 0.94 arcsec diameter fibres in three rings. HexPak is the first of two variable pitch integral field unit, feeding the Bench Spectrograph in a dual slit shared with the $\rm\nabla$Pak IFU \citep{Eigenbrot_2018}.  Of the time awarded 6 nights had good conditions, 7 nights had poor conditions, and 7 nights were not usable at all. A total of 43 galaxies were observed in the H$\rm\upalpha$ spectral region, which are reported here. 

These observations used the Bench Spectrograph configured for the R2 echelle (316 l/mm, blazed at 63.4 degrees). The echelle is used 11.5 deg off-Littrow with a grating incidence angle $\rm\upalpha=65.4$ deg. In order 8, selected via an order-blocking interference filter (X19), the spectra are centered at 675.7~nm and cover 654.9-694.3~nm on the STA1 2600x4000 pixel CCD. This device has 12~$\rm\mu$m pixels, but was used in a 2x2 binning mode to reduce read-noise. In this mode the system delivers 4.1 e- rms read-noise per 24~$\rm\mu$m super-pixel.

The Bench Spectrograph has a geometric demagnification factor of 0.356; with the echelle configuration, there is an additional anamorphic factor of 0.723 in the spectral dimension. Adopting the effective FWHM slit-width of a round aperture of diameter $D$ as $\rm\cos(30)\times D$, we expect the monochromatic image of the larger (smaller) fibres to be 2.8 (0.9) super-pixels in the absence of significant aberrations. With the delivered linear dispersion from the echelle of $0.019$~nm super-pixel$^{-1}$, we anticipate a spectral resolution $R=\lambda/\delta\lambda \sim 12850$ for the large fibres and, in the absence of optical aberrations, three times higher for the smaller fibres. However, the latter are significantly under-sampled, and they only yield $\rm\sim$50\% higher spectral resolution than the larger fibers due to aberrations. Since accurate knowledge of the instrumental resolution is very much of the essence for this analysis, the delivered resolution is measured in Section~\ref{sec:measurements}.

Typical H$\rm\upalpha$ spectroscopic observations consisted of a total of one hour of integration split between three, 20-minute exposures for cosmic-ray removal. There was no dithering between frames so that the nominal field coverage retains the interstitial fibre gaps. For our purposes of mapping to the MaNGA data with complete coverage this sampling was adequate. The rotator on the WIYN Instrument Adaptor System was always positioned to keep HexPak oriented in the same manner as the MaNGA IFU observations.

Calibration data consists of bias, dark, dome-flat, and thorium-argon arc-lamp exposures. There is significant bias and dark structure in STA1; 50-100 frames of each were taken over the course of a run and combined to minimize contributed noise in the standard object frame reduction process. Dome-flats and arc-lamp frames were taken at several exposures to provide adequate counts in the small fibres yet avoid saturation  in the large fibres. It is worth noting in what follows below that the arc-lamp light injection into the fibres does not follow the same light path as the dome-flats or sky, and as a consequence likely illuminates the fibres with a different \textit{f}-ratio.

Additional observations of seven galaxies in the H$\rm\upalpha$ sample were made in the second and third semesters in a second, medium resolution configuration sampling from [OIII]$\rm\lambda$500.7~nm through to the Mg~I triplet near 517~nm. These data and results therefrom are reported in a later paper. Here we specify the configuration as it has bearing on our sample selection. A 3300 l/mm volume-phase holographic grating was used in a first order Littrow configuration with grating angle close to 59.5 degrees; no blocking filter was required. This configuration provides a central wavelength of 522.2~nm and coverage between 508.4 to 534.2~nm. Spectral resolution was $\rm\sim20$\rm\% lower than the echelle observations in the H$\rm\upalpha$ region.

% Total targets observed: 54;
% 46 - HaNS only, 7 - HaNS + O3Mg,  1 - O3Mg only
% 17B: 5 nights:  4 + 2 x 1/2,  echelle HaNS
% 18A: 5 nights:  2 echelle HaNS, 2 O3Mg
% 18B: 10 nights: 4 echelle HaNS, 6 O3Mg

Standard IRAF\footnote{IRAF was distributed by the National Optical Astronomy Observatory, which was managed by the Association of Universities for Research in Astronomy (AURA) under a cooperative agreement with the National Science Foundation.} tools designed for image processing ({\tt ccdred} package) and the Bench Spectrograph spectral extraction  ({\tt hydra} package) were adapted and used. The only significant augmentations were for (i) handling the two fibre sizes in the HexPak array, similar to what is discussed by \cite{Eigenbrot_2018} for dealing with the five fibre sizes in $\rm\nabla$Pak, and (ii) for sky-subtraction, as discussed below.

\begin{figure*}
\centering
  \includegraphics[width=0.8\textwidth]{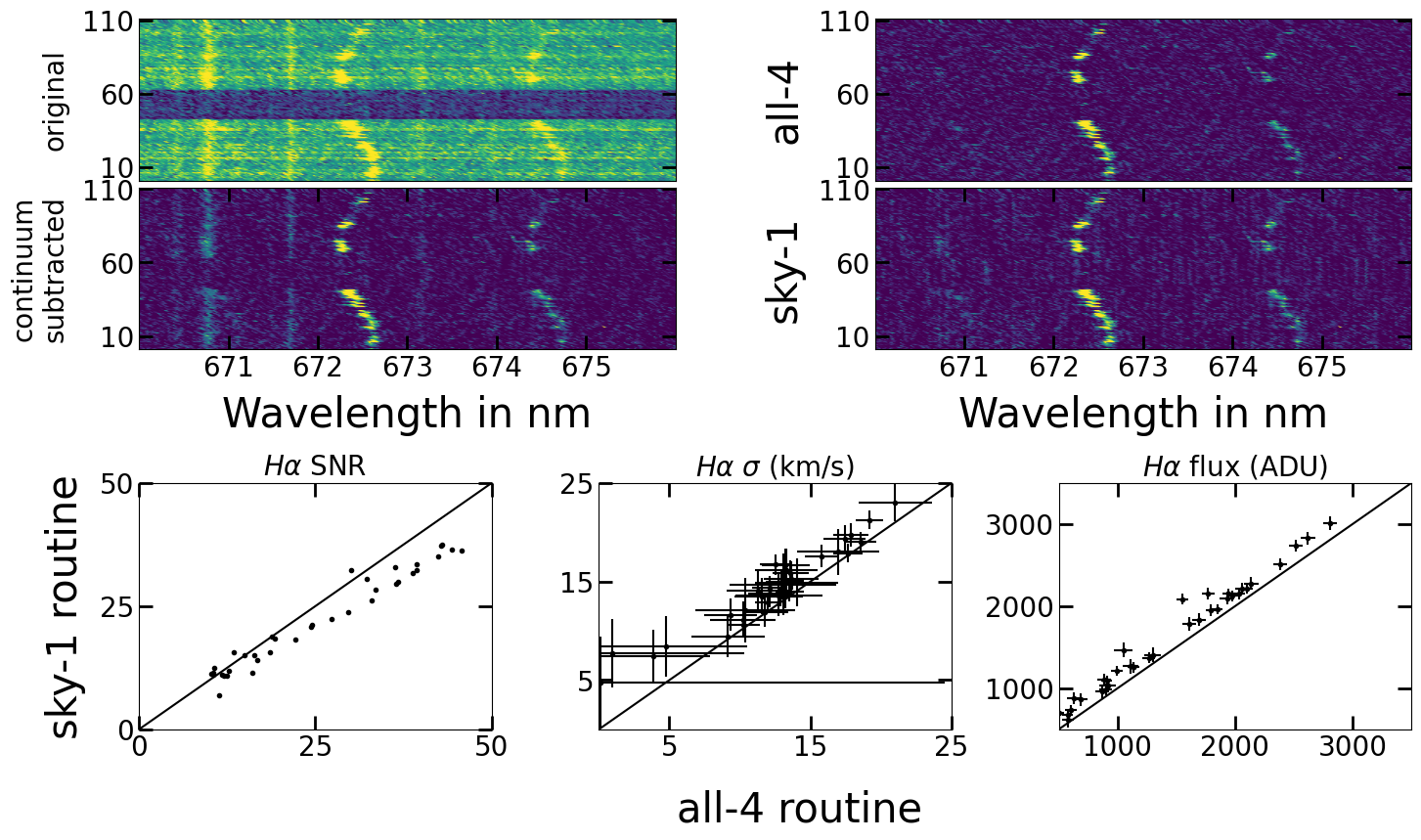}
\caption{Effects of sky-subtraction procedure on the measured SNR, line flux and line widths for galaxy 1-37018 in the H$\rm\upalpha$+[NII] region. The left top panel shows the reduced, wavelength calibrated, telluric-corrected, normalized spectra, while the left middle panel shows the same after continuum subtraction. Right top and middle panels show the continuum-subtracted spectra after sky subtraction through all-4 and sky-1 routines respectively. The HexPak fiber ID is on the y axis. The bottom three panels, left to right, compares SNR, line-width (\kms) and line-flux (instrumental units) measured for fibres with H$\rm\upalpha$ SNR$>10$ in the H$\rm\upalpha$ line emission measured from the all-4 sky-subtracted spectra (x-axis) and the sky-1 sky-subtracted spectra (y-axis).}
% higher SNR (SNR$\rm\geq$10) fibres shows that sky spectra estimated by fitting a 4th order polynomial to all large fibres over-subtracts source-flux and under-estimated the line-width even though it improves the SNR at least at higher SNR compared to sky spectra generated by fitting 1st order polynomial to large sky fibres.}
\label{fig:skysub}
\end{figure*}

\subsubsection{Sky subtraction}
\label{sec:skysub}

HexPak has 7 large and 2 small sky fibres. We implemented a custom IRAF routine for independent sky subtraction of large and small HexPak object fibres using these sky fibres based on concepts illustrated in \cite{Bershady05}. This subtraction operates on wavelength-calibrated, field-flattened and rectified spectra; in IRAF these are referred to as `ms' files, while in MaNGA they are referred to as 'row-stacked spectra.' The data format arranges the spectra in order as they appear in the fibre pseudo-slit, with every wavelength channel aligned in one data index. 

The subtraction routine begins by fitting a polynomial to the continuum of every fibre (including sky fibres) and subtracting it out. The continuum-subtracted spectra are used for source emission line analysis. The continuum fitting uses a $\rm\pm3\upsigma$ clipping to remove source and sky emission lines and any remaining detector artefacts from the estimate of the continuum. In the case of the echelle data a third-order Legendre polynomial was adequate to represent the continuum. The continuum spectral fits are saved for spatial registration purposes (Section~\ref{sec:registration}), which for this purpose have the mean sky-fibre continuum spectra (large and small, separately) subtracted.

Given the few small sky-fibres, the two continuum-subtracted sky spectra were simply averaged and then subtracted from the small object fibres, leaving the final, continuum- and sky-line subtracted spectra for final emission-line analysis. The large sky fibres trace uniformly (but sparsely) along the full pseudo-slit. These were fit separately in each wavelength channel with a low order function. The function is then interpolated along the pseudo-slit and subtracted from the object fibres for final emission-line analysis. The purpose of a non-constant function is to model the variation in the sky-line profiles due to changes in aberrations and sampling along the fibre pseudo-slit. Higher-order functions (order three to five) are desirable but are not well constrained with the small number of sky fibres. In the end, we reduced to order one (a constant, essentially the same as a clipped mean) to minimize residual structure in the line-free regions; order three minimizes residuals in bright lines, but adds substantially more structure in line-free regions. The line-subtraction is robust but still adds visually detectable coherent noise to the 2D spectra since only 7 fibres are being used to determine the sky-line level. The coherent noise is at a low level, and it is incoherent with respect to the source line-emission that is significantly doppler-shifted from fibre to fibre. Hereafter we refer to this as the `sky-1' routine.

Because the data is at such high dispersion, most of the fibre channels at any given wavelength are free from source line-emission even at wavelengths where, e.g., H$\rm\upalpha$ and other nebular lines are present. This offers the potential opportunity to reduce the noise introduced in sky-line subtraction by using \textit{all} of the large fibers, with sigma clipping, to determine the sky-line level which, in contrast will be nearly constant in all fibres. The small fibres are treated as before in the sky-1 routine. With more large fibres, we are able to fit and then subtract a 4th order polynomial with $\rm\upsigma$-clipping as before for the continuum, here to remove sky-line emission. We refer to this as the `all-4' routine. Visually, the sky-subtraction appears to be vastly improved in terms of reducing coherent noise from spectral channel to channel, but some over-subtraction of source flux is apparent in some fibres despite attempts to optimize the $\rm\upsigma$-clipping. Since over-subtraction could systematically clip source emission-line wings, this is a concern. Therefore we compared Gaussian fitting results (described below) between the two sky-subtraction routines. Figure~\ref{fig:skysub} illustrates, for $\rm\sim$37 fibres with signal-to-noise (SNR) $\rm\geq10$ for galaxy 1-37018, the difference between the all-4 and sky-1 routines. We find that the all-4 routine over-subtracts source-flux and under-estimates line-width even though it improves SNR, albeit only at higher SNR where it is a less-useful gain. As a consequence, we adopted the sky-1 routine for our analysis.

% Notes for later addition: The redshift range for the targets puts Ha+[NII] in a very clean portion of the sky spectrum. The [SII] lines are sufficiently redshifted however to put them in a region of relatively strong sky line emission as well as the beginning of the Fraunhofer B band (Telluric O2). We corrected for the absorption in the usual way using spectra of hot stars with their continuum fit and normalized.

\section{Measurements}
\label{sec:measurements}

For consistency with previous MaNGA analysis and best practice we adopt {\tt pPXF} \citep{Cappellari04, Cappellari_2017} to measure the emission-line centers, widths and fluxes of calibration arc lines, sky emission lines as well as galaxy nebular emission. In all cases we fit a single-component Gaussian line profile to each line. In the case of astrophysical nebular lines including H$\rm\upalpha$, [NII]$\rm\lambda\lambda$ 654.8,658.4~nm, and [SII]$\rm\lambda\lambda$671.7,673.1~nm we use the {\tt ppxf} package with a single velocity and line-width for all lines, while allowing flux to vary, with the exception of the 3:1 ratio for the [NII] lines. However, for arc and sky emission lines, a modified version of the {\tt emission\_line} routine of {\tt ppxf\_util.py} is used for line wavelengths obtained from NOAO arc lamp catalogue within the observed wavelength range. Again using {\tt ppxf}, we individually fitted Gaussian profile to extract instrumental dispersion to $\sim$35 arc lines separately which sparsely sample the wavelength range with velocity constraints of $\pm$20km/s.

\subsection{Measurement of the Bench Spectrograph LSF from arc-lamps}
\label{sec:arcs}

Individual arc-lamp line wavelengths were fed into {\tt pPXF} along with the arc-lamp spectra, given an initial velocity and line-width estimate. The {\tt pPXF} widths, $\rm\upomega_{arc}$, measured at the discrete arc-line wavelengths, are interpolated over wavelength using a 5th order Legendre polynomial. The polynomial degree reflects the shape of the focus curve for the Bench Spectrograph dioptric camera, which has a shape of a (sometimes tilted) ``Mexican hat." This interpolation in $\rm\upomega_{arc}$ is shown in Figure~\ref{fig:arclinewidth} in pixels and equivalent 1-$\rm\upsigma$ instrumental velocity resolution. Variations of instrumental line-width as a function of wavelength of this order ($\rm\pm20$\rm\%) are typical for the Bench Spectrograph. The characteristic instrumental resolution for the larger fibres is $\rm\sim$10 km s$^{-1}$, or R$\rm\sim$12,740, very close to our expectations from Section~\ref{sec:hex}. We note that given our sampling is close to the critical value, the impact of the finite pixel size on our Gaussian width estimates should not be ignored \citep{Robertson_2017, Law_2021}. We use the default value \texttt{pixel=True} in the creation of the gas emission line templates in \texttt{ppxf\_util.emission\_lines}. By using {\tt pPXF} to estimate these widths, in the parlance of \cite{Law_2021}, we are measuring and reporting pre-pixelized estimates of the instrumental Gaussian line-width. Given our sampling this is likely to be about 3\% smaller than the post-pixelized values that would be estimated from simply fitting Gaussian functions evaluated at the pixel centers. The smaller fibres have characteristic instrumental resolution of $\rm\sim6.5$ km s$^{-1}$, or R$\rm\sim$19,600. These too are pre-pixeled Gaussian line-widths; their post-pixelized counterparts would be significantly larger. As an aside, we note the fact that the pre-pixelized LSF for the small fibres does not scale with the geometric size indicates that there are significant contributions from optical aberrations at the physical scale of the reimaged fibre FWHM of $\rm\sim22~\mu$m at the detector; we estimate the effective aberrations $\rm\upomega_{\rm abb}\sim16\mu$m, again as a pre-pixelized value.

\subsection{Measurement of the Bench Spectrograph LSF from sky lines}
\label{sec:sky}

\begin{figure}
\includegraphics[width=\columnwidth]{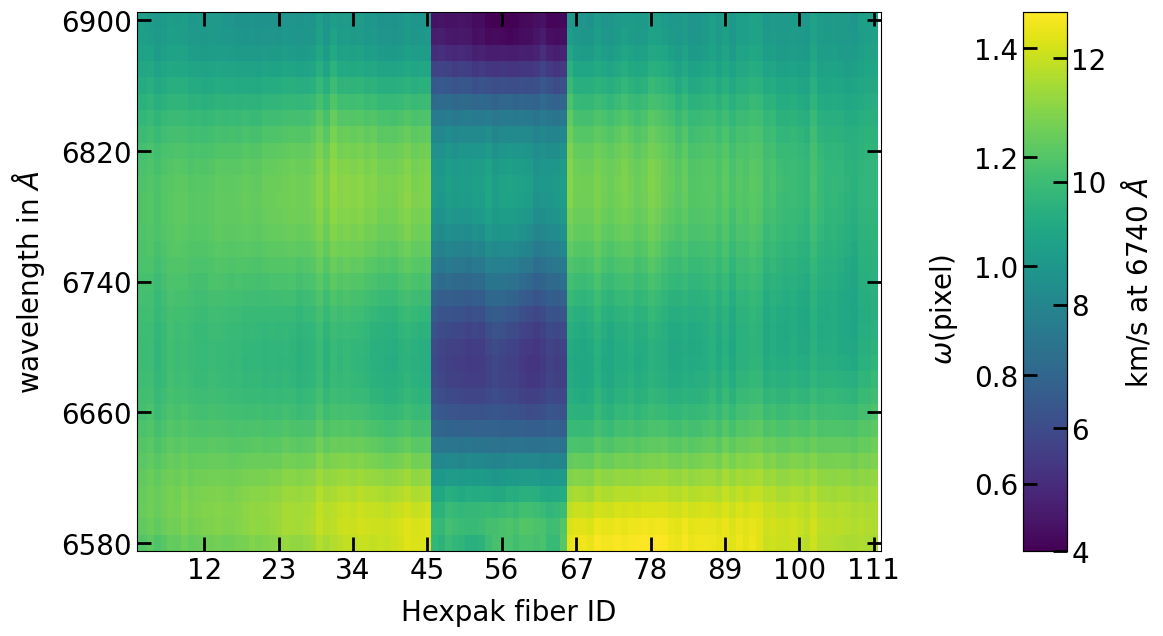} 
\caption{Distribution of arc line-widths (Gaussian width, $\rm\upomega$, in pixels) across different HexPak fibres and the observed wavelength range of 658-690~nm. Arc-lamp line-widths, measured using {\tt pPXF} are interpolated with a 5th order Legendre polynomial in wavelength and rendered here at steps of 1~nm, i.e., at an interval of 20 pixels. The color-bar is also referenced with the velocity equivalent of the arc line widths ($\rm\upsigma$) at the median wavelength of 674~nm. The  fibres in the middle (fibre ID 45 to 65) with small instrumental line-widths are the 0.94\arcsec\ fibres.}
\label{fig:arclinewidth}
\end{figure}

% MAB: Maybe we should just go back to showingomega_arc / omega_sky ?

\begin{figure}
\centering
\includegraphics[width=\columnwidth]{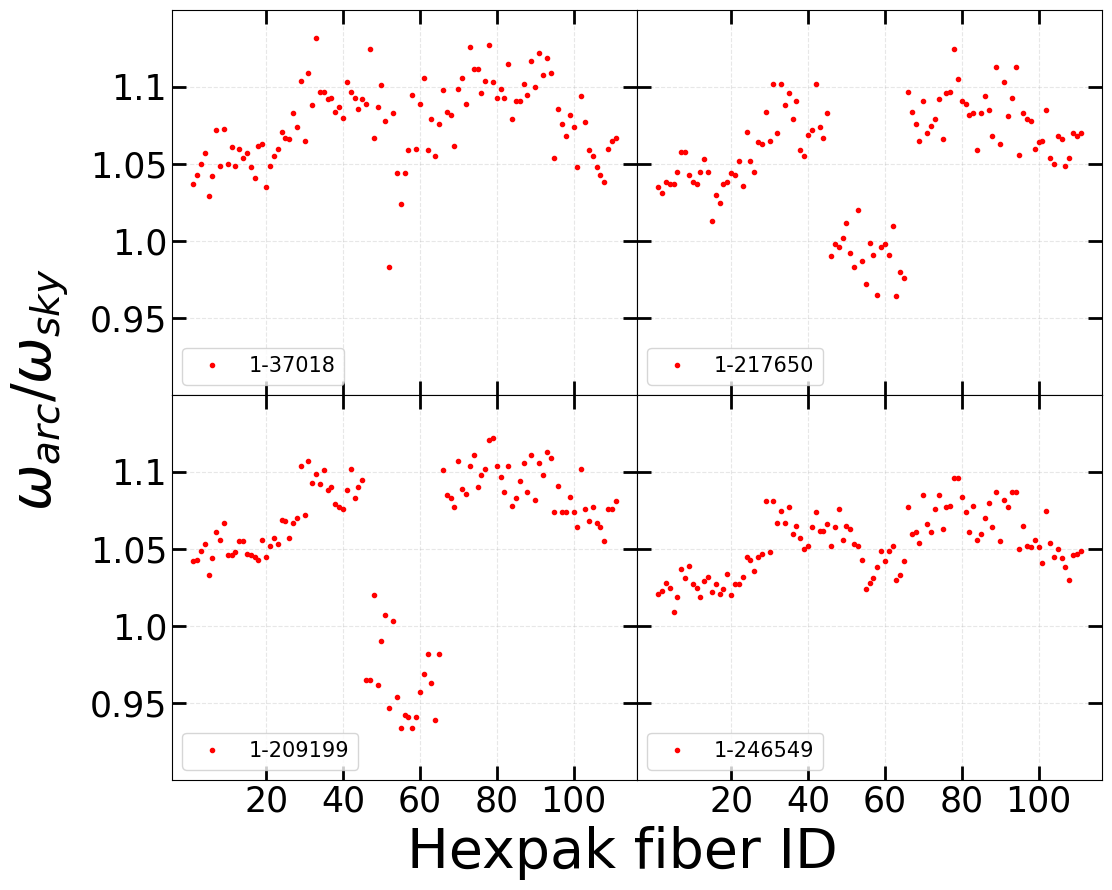}
\caption{Ratio of HexPak instrumental width measured using LSF estimates from arc and sky lines ($\rm\upomega_{arc}$ and $\rm\upomega_{sky}$) at H$\upalpha$ wavelengths. Panels show different galaxies observed on different nights and runs. Large and small fibres exhibit differences up to -6\% (small fiber) and+12\% (large fibers). The outer large (2.8\arcsec) fibres tends to have smaller $\rm\upomega_{arc}$ compared to central large fibres.}
\label{fig:sky_arc_comp}
 \end{figure}

\begin{figure*}
\centering
\includegraphics[width=0.9\textwidth]{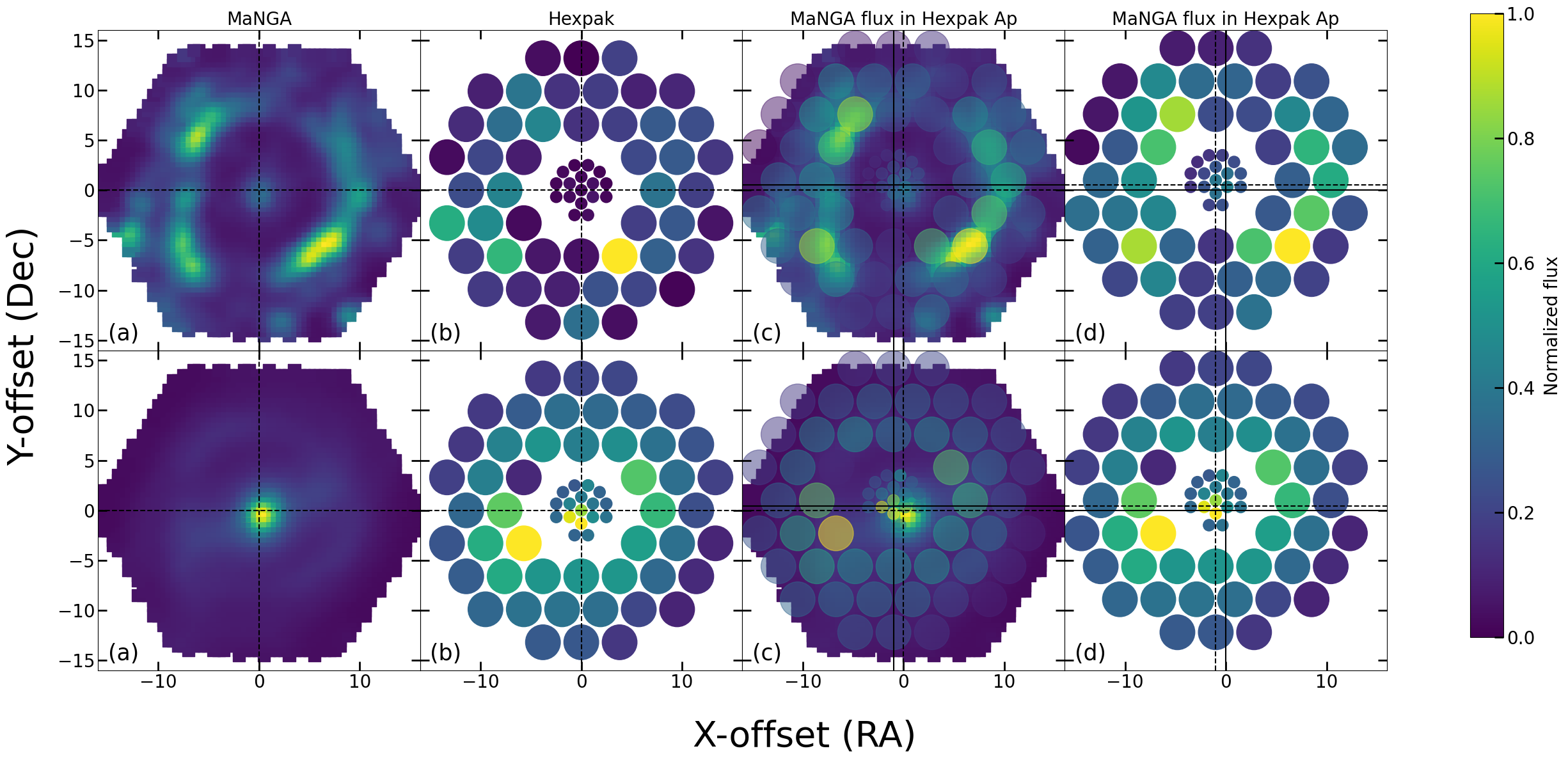}   
\caption{Typical X (RA) and Y (Dec) offsets between MaNGA and HexPak IFU pointings determined via cross-correlation of H$\rm\upalpha$ line flux (top panels) and stellar continuum (bottom panels) for the galaxy 1-37211. Axes in each panel are sky-offsets in arcsec. Continuum offsets are (-0.4,1.0) arcsec in (X,Y), and (-0.5,1) for H$\rm\upalpha$. Panels from left to right shows (a) MaNGA measured flux, (b) HexPak measured flux, (c) MaNGA measured flux overlaid with MaNGA measured flux within the HexPak fibre footprint, and (d) MaNGA measured flux within the HexPak fibre footprint. The dashed horizontal and vertical lines in each panel represent field centers for the MaNGA datacubes (columns 1 and 3) and HexPak array (columns 2 and 4).
Solid lines show the center offset between HexPak and MaNGA (column 3) and MaNGA and HexPak (column 4) after registration.}
\label{fig:ccf}
\end{figure*}

Sky-line widths are measured in the continuum-subtracted object spectra saved from the initial sky-subtraction stage described in Section~\ref{sec:skysub}, again individually for all fibres. There are fifteen sky lines in the observed wavelength range out of which only eleven were fit with {\tt pPXF} to derive $\rm\upomega_{sky}$. These eleven lines were selected to ensure both arc and sky lines have wavelength overlap for fair comparison of $\rm\upomega_{arc}$ to $\rm\upomega_{sky}$.

We find, consistently, that for the large HexPak fibres the sky lines yield smaller instrumental line-widths ($\rm\upomega_{sky}<\upomega_{arc}$), while the opposite holds for the smaller fibres ($\rm\upomega_{sky}>\upomega_{arc}$). As demonstrated in figure \ref{fig:sky_arc_comp}, the differences are significant (e.g., the means and the error in the means of $\rm\upomega_{arc}/\upomega_{sky}$ for large and small fibres are  1.08$\pm$0.002 and 0.97$\pm$0.006 respectively for the galaxy 1-209199), but because the instrumental line-widths are so small, the impact of these differences are very modest on corrected astrophysical line-widths that are likely for ionized gas. To illustrate this, we make the following simple calculation. Assuming Gaussian line profiles, we adopt the observed line-width $\rm\upsigma_{\rm obs}=22.4$~\kms\ so that the average corrected astrophysical line-width is $\rm\upsigma=20$~\kms. This value was chosen because it agrees with the mean value of our HexPak corrected measurements, as shown below. The uncertainty associated with this corrected line-width due to the variation between sky and arc LSFs is $\pm0.3$~\kms. Convolving this astrophysical line-width and uncertainty with the nominal MaNGA LSF of 67.6~\kms\ yields an observed line-width for MaNGA of $70.5\pm0.1$~\kms. If this uncertainty were inferred as an uncertainty in the MaNGA LSF that would lead to commensurate MaNGA LSF uncertainty of only 0.14\%.

There is also a visible trend in the LSF differences between arc and sky lines for the large fibres within a galaxy. The large  fibres closer to the slit center demonstrate larger instrumental LSF differences compared to the edge fibres. This \textit{might} be explained by the vignetting profile of the Bench Spectrograph: The redesigned collimator \citep{Bershady_2008, Knezek_2010} is optimized for an f/5 injection beam, but sized for f/4 at the field edge (edge of the slit). Given the uncertainty of injection speed from the arc lamps, the arcs \textit{may} have a faster output beam from the fibres than the sky. At the center of the slit where there is the least vignetting, more of the light entering at larger angles (in a faster beam) will get through and lead to systematically larger aberrations, and hence larger arc line-widths compared to the sky lines. This model does not explain the different behavior between the large and small fibres, but this shortfall does not impact our remaining analysis.

We performed the analysis in Section~\ref{sec:analysis} after making a correction for the difference between $\rm\upomega_{arc}$ and $\rm\upomega_{sky}$. As in \cite{Law_2021}, because the sky-lines do not sample wavelengths as well as the arc-lines, the correction will be a suitable approximation. In the case of the HexPak echelle data, there are two clusters of sky lines, one at the red end and the other at the blue end of the wavelength range. The bluer cluster of sky-lines shows a slightly larger offset in $\rm\upomega$ between sky and arc than the redder cluster of sky lines. We fit a linear function in wavelength to the difference $\rm\upomega_{arc}-\upomega_{sky}$ to the full set of sky lines and apply this linear function to the Legendre polynomials shown in Figure~\ref{fig:arclinewidth}. As a consequence of making this correction we estimate any systematics in the corrected HexPak ionized gas line-width introduce systematics in our estimate of the MaNGA LSF well below 0.1\%.

\begin{figure}
\centering
\includegraphics[width=0.75\columnwidth]{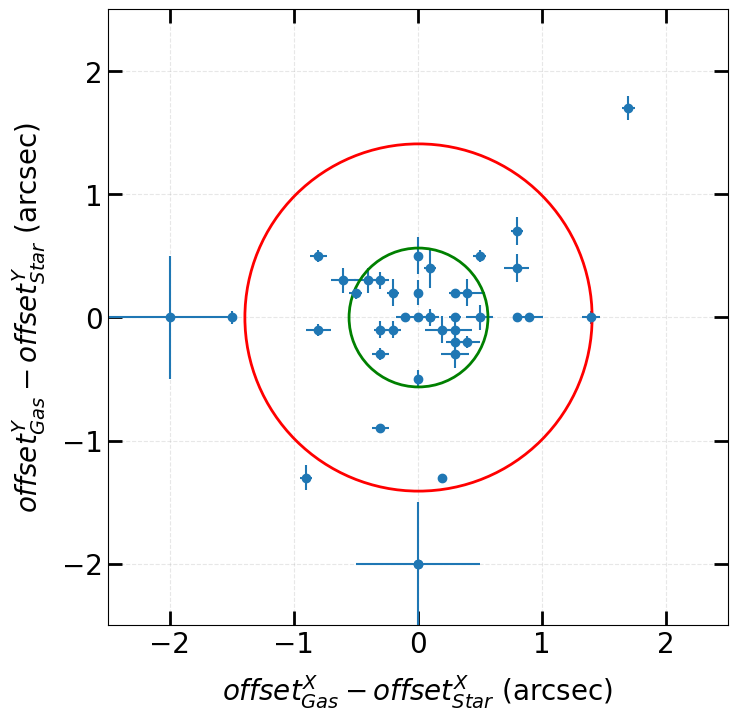}   
\caption{Differences between HexPak and MaNGA IFU spatial registration when estimated via cross-correlation of the H$\rm\upalpha$ flux distribution versus the stellar continuum distribution. Here X is RA and Y is Dec. Each point is a galaxy in our sample. The red and green circles represent footprints of the large and small fibers, respectively.}
\label{fig:ccf_summary}
\end{figure}

\subsection{Spatial registration}
\label{sec:registration}

We used a simple cross-correlation method to find the relative positions of the HexPak and MaNGA IFUs on the target galaxies, assuming no angular misalignment. Given the higher- resolution MaNGA data-cubes, this cross correlation sampled the MaNGA data-cubes given the HexPak fibre footprint in a process that is otherwise similar to what is described in \citep{Bershady05} for the SparsePak IFU. This process was done independently for the stellar continuum and the H$\rm\upalpha$ line-emission as shown in figure \ref{fig:ccf} to gauge the uncertainty in our estimates of the relative positioning between MaNGA and HexPak IFUs. For this example, the offsets are small and consistent between gas and star, as indicated by the dashed and solid horizontal and vertical lines in the figure. For the sample overall, as shown in Figure \ref{fig:ccf_summary}, these two tracers yielded comparable results, with spatial offset measurements differing by $<$0.5\arcsec for about 70\% of the sample. In 35 out 40 galaxies the offset is within the large fiber footprint, and hence overall the offsets are small. The mean and standard deviation of the offset differences are 0.03$\rm\pm$0.67\arcsec and 0.07$\rm\pm$0.65\arcsec along X and Y directions, respectively. In terms of a radial offset differences between measurements from gas and stars, we find 90\% of the galaxies have difference $<$ 0.5\arcsec, with mean and standard deviation of 0.08$\rm\pm0$.34\arcsec.  This indicates that the uncertainties in our spatial registration of the two IFU maps are well within their fibre footprints of 2.81 \arcsec in diameter, and even the sub-sampled (0.5 arcsec) MaNGA datacubes.

\section{Analysis}
\label{sec:analysis}

\subsection{Measurement biases in the corrected, Gaussian line-widths}
\label{sec:biases}

In the presence of measurement error, two biases manifest in the estimation of Gaussian line-widths corrected for the effect of instrumental broadening: (i) positivity bias and (ii) survival bias. 

Positivity bias comes from fitting a function with positive-definite parameters to noisy data. In our case, this is a Gaussian function with a dispersion parameter $\rm\upsigma\geq0$. The bias is not specific to the fitting routine (e.g., {\tt pPXF}), but is a generic attribute of functional fitting with bounded parameters. As the uncertainty in the dispersion parameter estimate -- due to errors in the data being fit --  become comparable to, or greater in magnitude than, the actual line width, the distribution of the fitted dispersion parameters is biased, statistically, towards larger values than the underlying (actual) line width.

Survival bias comes from the numerical evaluation of the correction, in quadrature, of the measured dispersion parameter, $\rm\upsigma_{\rm obs}$, and the instrumental line-width, $\rm\upomega$, to estimate the astrophysical line-width, again: $\rm\upsigma = \sqrt{\upsigma_{\rm obs}^2-\upomega^2}$. In the presence of measurements errors on both $\rm\upsigma_{\rm obs}$ and $\rm\upomega$, the argument of the square-root can be negative in some measurement instances; the chance of this happening increases (up to $\rm\sim$50\%) as the combined measurement errors in $\rm\upsigma_{\rm obs}$ and $\rm\upomega$ become comparable to, or larger than $\rm\upsigma$, i.e., at low SNR and small $\rm\upsigma/\upomega$. In numerical analysis, imaginary values typically are censored from statistical computations, which in effect truncates the error distribution and systematically biases the distribution of estimated $\rm\upsigma$ to larger values. 

Both biases act to increase the corrected line-width above the intrinsic value, and these biases \textit{increase} with \textit{decreasing} SNR and, in the case of survial bias, with \textit{lower} spectral resolution (\textit{larger} $\rm\upomega$). The effects of these biases in MaNGA data are well known: \cite{Westfall_2019} discuss and simulate the `positivity boundary bias' (what we call positivity bias here) in their Section 7.5.2, while \cite{Law_2021} simulate the impact of both biases in their Section 4.3. Both positivity and survival biases increase significantly the estimated MaNGA dispersion in the regime of the data evaluated in our study, well below the instrumental resolution. For similar reasons \cite{Law_2021,Law_2021b} restrict their analysis of MaNGA gas line-widths to SNR$>$50. Nonetheless, the corrections can become significant, and depend critically on the adopted functional form of the error distribution, as noted by both \cite{Westfall_2019} and \cite{Law_2021}, and as we discuss below.

In contrast, since the HexPak instrumental dispersion (the line-width $\rm\upomega$) is lower than the expected astrophysical dispersions,  $\rm\upsigma$, in our data, we expect both the positivity and survival biases to be negligible for SNR $\rm\geq10$; see Figure 20 in \cite{Westfall_2019} and Figure 15 in \cite{Law_2021}, respectively. We restrict our consideration of HexPak line-widths to this SNR regime.

One way to ameliorate the impact of survival bias on MaNGA data\footnote{We thank M. Blanton for pointing this out.} is to work directly with the statistical distribution of $\rm\upsigma^2$. As long as the mean or median of the \textit{distribution} of $\rm\upsigma^2$ remains positive, the square-root of the mean or median becomes an unbiased estimator of $\rm\upsigma$. In general, the mean of the square-root is not equal to the square-root of the mean, so the median statistic is preferred. For this reason we propose the square-root of the median of the measurement distribution of $\rm\upsigma_{\rm obs}^2-\upomega^2$ as an estimator of (the median) $\rm\upsigma$. In the following section we use new simulations to ascertain if this statistic is preferred over the corrected mean values described in \cite{Law_2021}.

\begin{figure*}
\centering
\includegraphics[width=0.8\textwidth]{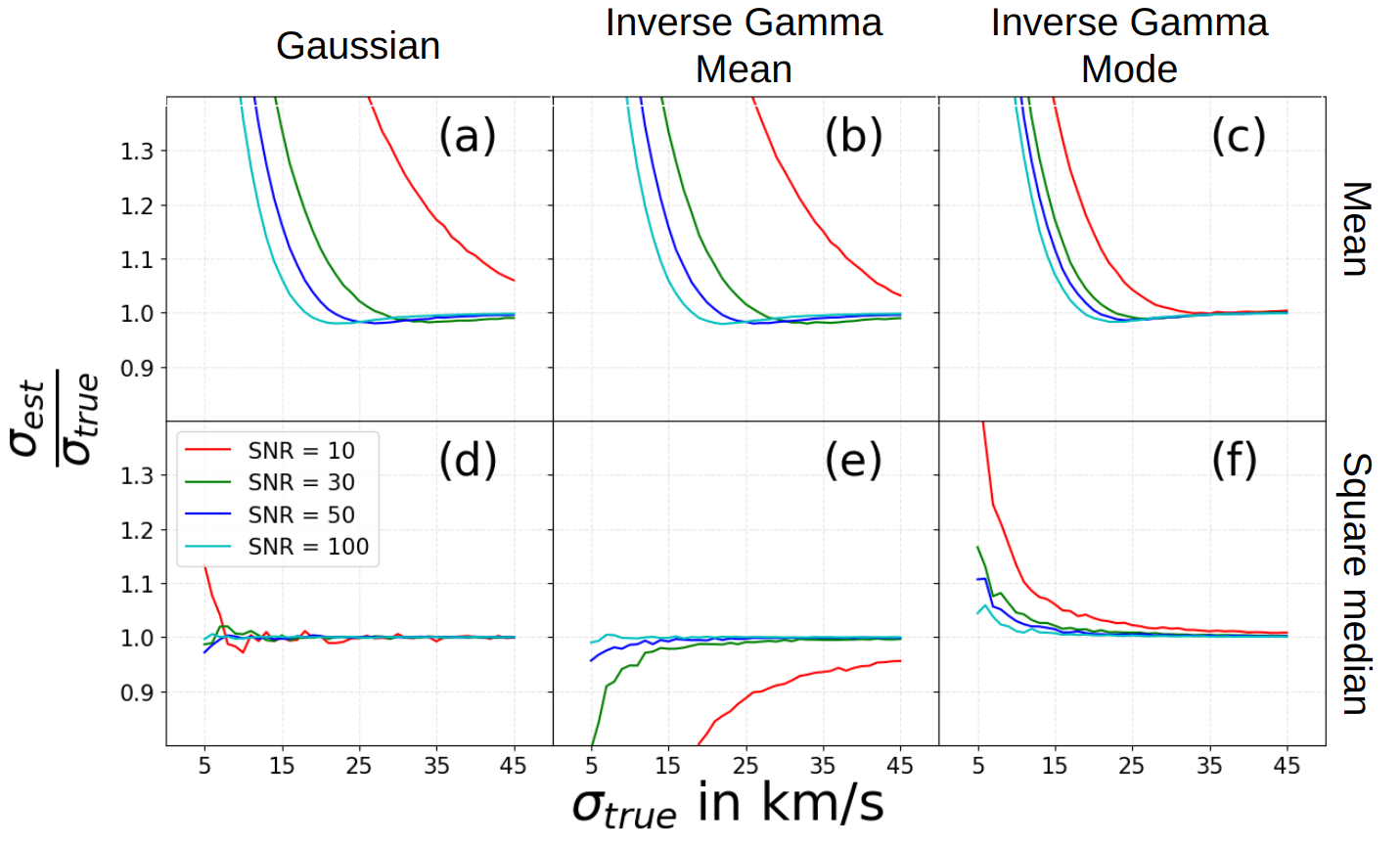}   
\caption{Simulations showing the percentage of bias in the recovered velocity dispersion for different SNR, different estimation methods, and different error distribution functions as a function of velocity dispersion. The ratio of recovered to model (``true") velocity dispersion is plotted against the model dispersion value. The top panels (a,b,c) show the fractional bias introduced by mean estimation while the bottom panels (d,e,f) shows the same for square median estimation. Panels (a) and (d) show the performance adopt a Gaussian error distribution while the remaining panels adopt the inverse gamma function for the error distribution. The middle panels (b,e) and right panels (c,f) are computed with mean and mode formalization of the inverse gamma distribution.}
\label{fig:bias_sim}
\end{figure*}

\begin{figure}
\centering
\includegraphics[width=\columnwidth]{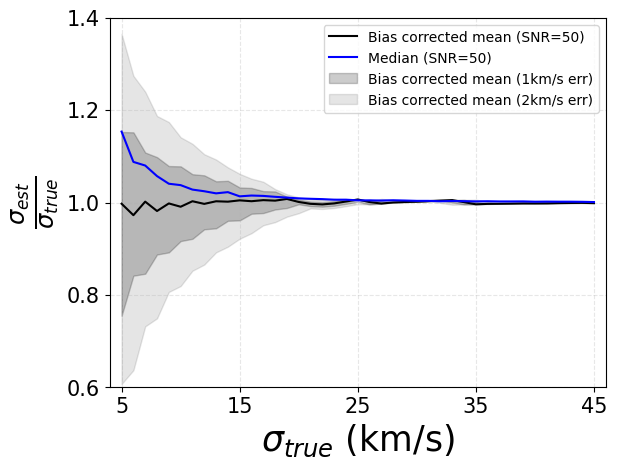}   
\caption{Effects of errors in estimating the intrinsic dispersion to compute the bias correction factor. The simulation is performed for the inverse gamma error distribution with the mode formalism. The solid black line represents the bias-corrected mean of the LSF-corrected dispersion, assuming no error in estimating the intrinsic dispersion, while the dark and light-grey shaded regions bound areas with limiting errors on estimating the intrinsic dispersion of 1 and 2~\kms\, respectively, in computing the bias correction. The blue line is for the square-root of the median of the squared difference between the measured dispersion and the LSF (refer to panel (f) in Figure~\ref{fig:bias_sim}), for which there is no correction applied, and hence this measure does not depend on errors in the estimated intrinsic dispersion.
}
\label{fig:guess_sim}
\end{figure}

\subsection{Simulations of positivity and survival bias in the corrected Gaussian line-widths}
\label{sec:simulations}

We evaluated the bias in recovered velocity dispersion as a function of intrinsic dispersion and SNR via a Monte Carlo simulation that used two different error distributions: a Gaussian function and an inverse gamma function (hereafter IGF). While the IGF may not be an immediately intuitive choice for the error distribution, in Bayesian statistics the error distribution of the width parameter for a Normal distribution, $\rm\upsigma$ is indeed drawn from the IGF \citep[e.g.,][]{MacKay_2003}. Specifically, the probability density distribution of $\rm\upsigma^2$ is ${\rm P}(\upsigma^2) = \upbeta^\upalpha \ \upGamma(\upalpha)^{-1} \ \upsigma^{-2(\upalpha+1)} \ \exp(-\upbeta/\upsigma^2)$ where $\rm\upalpha$ and $\rm\upbeta$ are related to attributes of the distribution and $\rm\upGamma$ is the gamma function. By fitting simulated Gaussian line profiles generated with random noise, it is straight forward to show that the measurement distribution appears to be well characterized by a Gaussian in the limit when the first moment of the distribution (e.g., the mean or mode of $\rm\upsigma$) is larger than the second moment (e.g., the square-root of the distribution variance); in this case the mean and mode are nearly equivalent. Indeed, this is the expected limiting behavior of the IGF. However, when the second moment of the distribution is of order, or greater than the first moment, the distribution is asymmetric about the mode, which is also well represented by the IGF. In this limit the mean value of the distribution departs significantly from the mode, and the distribution is not well characterized by a Gaussian.

Unfortunately, there is some ambiguity in connecting the IGF functional parameters $\rm\upalpha$ and $\rm\upbeta$ to what might be observed in terms of unbiased estimators. For example, in the case of a Gaussian error distribution of the Gaussian line-width parameter $\rm\upsigma$, the distribution of the observed values of $\rm\upsigma$ in the presence of measurement-error characterized by a variance $\rm\upepsilon^2$ is expected to have a mean of $\rm\upsigma$ and full-width half-maximum of $2.355\upepsilon$. For the IGF, however, while the variance of the distribution can be associated with $\rm\upepsilon^2$, it is unclear if the IGF mode or mean (or some other statistic) is most appropriate to associate with $\rm\upsigma$. The two choices of mean or mode pose the easiest analytic forms for solving for the IGF parameters $\rm\upalpha$ and $\rm\upbeta$; from our simulations they appear likely to bracket a physical description relevant to the sampling parameters and astrophysical distributions of our data.

Hence here we consider these three different distributions that describe measurements in the presence of noise: (a) Gaussian; (b) IGF with a mean formalism, where the distribution mean $\rm\mu = \upbeta/(\upalpha-1)$; and (c) IGF with a mode formalism, where the distribution mode $m = \upbeta/(\upalpha+1)$. The latter was adopted by \cite{Law_2021}. Since the IGF can have significant skew, these choices impact the outcome of our simulations.

In our simulations we considered cases with SNR values of 10, 30, 50 and 100 as well all intrinsic dispersion values of $5 < \upsigma_{\rm true} < 45$ km s$^{-1}$. For each combination of SNR and intrinsic dispersion, we computed the nominal dispersion ($\rm\upsigma_{\rm nom}$) by convolving the intrinsic dispersion with nominal width of the LSF:
\begin{equation}
    \upsigma_{\rm nom} = \sqrt{\upsigma_{\rm true}^2 + \upomega_{\rm LSF}^2}.
    \label{eq:sigma_nom}
\end{equation}
In the context of the previous section we can equate $\rm\upsigma_{\rm nom} = \upsigma_{\rm obs}$ and $\rm\upsigma_{\rm true} = \upsigma$. 

For each SNR and $\rm\upsigma_{\rm true}$ we then created an observed distribution of $10^5$ samples of $\rm\upsigma_{\rm nom}$ with a distribution width ($\rm\upepsilon$) given by the relationship suitable for MaNGA data obtained from Figure 14 in \cite{Law_2021}:
\begin{equation}
    \upepsilon = \upsigma_{\rm nom} \times 10^{-1.08\times \log({\rm SNR})+0.24}.
        \label{eq:errors}
\end{equation}
For the Gaussian error distribution % (a) and the IGF error distribution (b), t
the observed distribution of $\rm\upsigma_{\rm nom}$ has a mean equal to the nominal dispersion given in equation~\ref{eq:sigma_nom}. For the IGF error distribution the value in equation~\ref{eq:sigma_nom} is equated either with the distribution mean (b) or mode (c).

Independently, we created an observed distribution of the LSF, $\rm\upomega$, with the same number of samples as $\rm\upsigma_{\rm nom}$. Following \cite{Law_2021}, the distribution had a mean value of $\rm\bar{\upomega} = 67.6$ km s$^{-1}$ and a Gaussian error distribution standard deviation, $\rm\upepsilon_\upomega$, such that $\rm\upepsilon_\upomega/\bar{\upomega} = 0.03$. These values well represent the expected distribution for the MaNGA LSF estimates. We then subtracted the square of each element in the LSF distribution (a randomized list) from the square of the corresponding element in the $\rm\upsigma_{\rm nom}$ distribution (an independently randomized list) to obtain the recovered dispersion distribution, $\rm\upsigma_{\rm est}^2$.

\begin{figure*}
\centering
\includegraphics[width=0.95\textwidth]{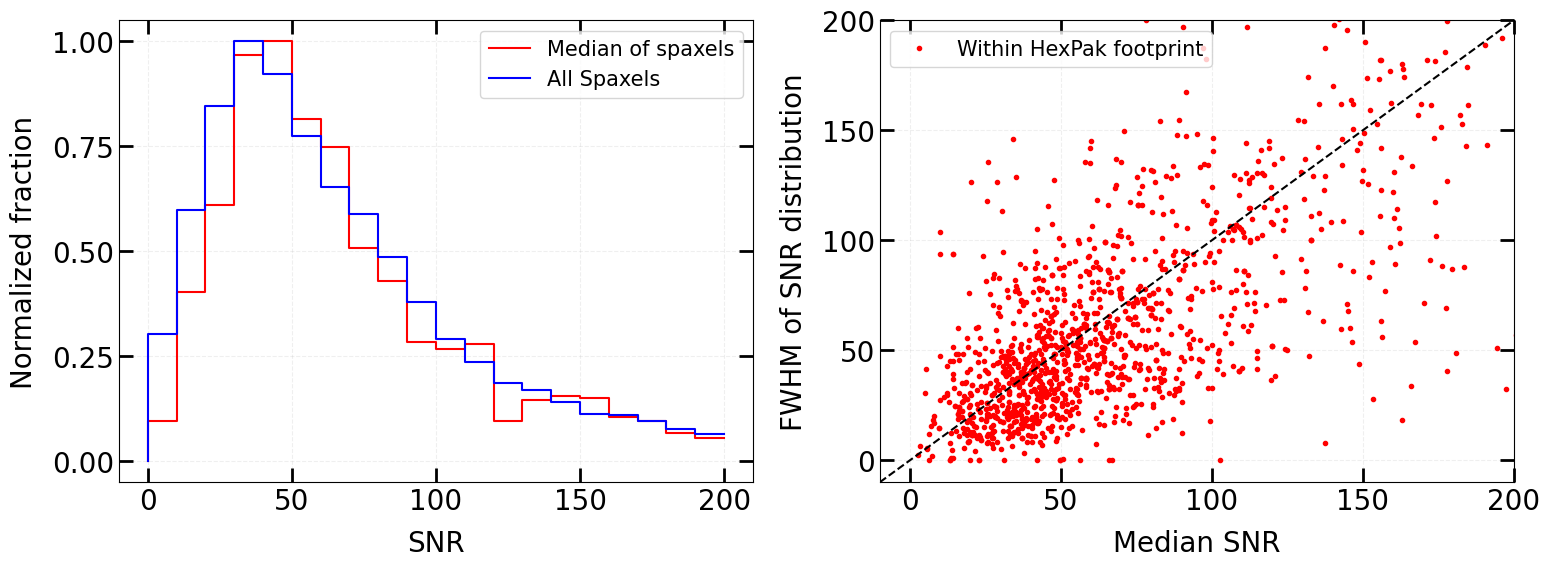}
\caption{SNR distributions of MaNGA spaxels within HexPak footprints at radii between 4 and 15\arcsec and HexPak SNR$>$10. Left: Histogram of all spaxels  (blue) and the median (per footprint) of the same spaxels (red) . Right: FWHM versus median of MaNGA SNR distribution within each HexPak footprint.}
\label{fig:snr_dist}
\end{figure*}

\begin{figure}
\centering
\includegraphics[width=\columnwidth]{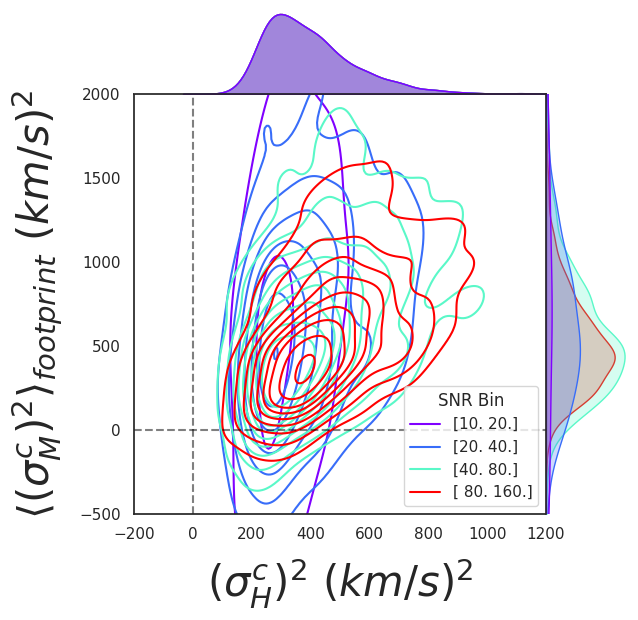}   
\caption{Distribution contours of LSF-corrected H$\upalpha$ dispersions for the median MaNGA value for spaxels within HexPak footprints versus HexPak where the HexPak fibers have SNR $>$ 10. Contour colors represent logarithmically increasing SNR intervals as defined in the key. Marginal histograms have identical SNR intervals. } 
\label{fig:sigma_hp_manga}
\end{figure}

\begin{figure*}
\centering
\includegraphics[width=0.9\textwidth]{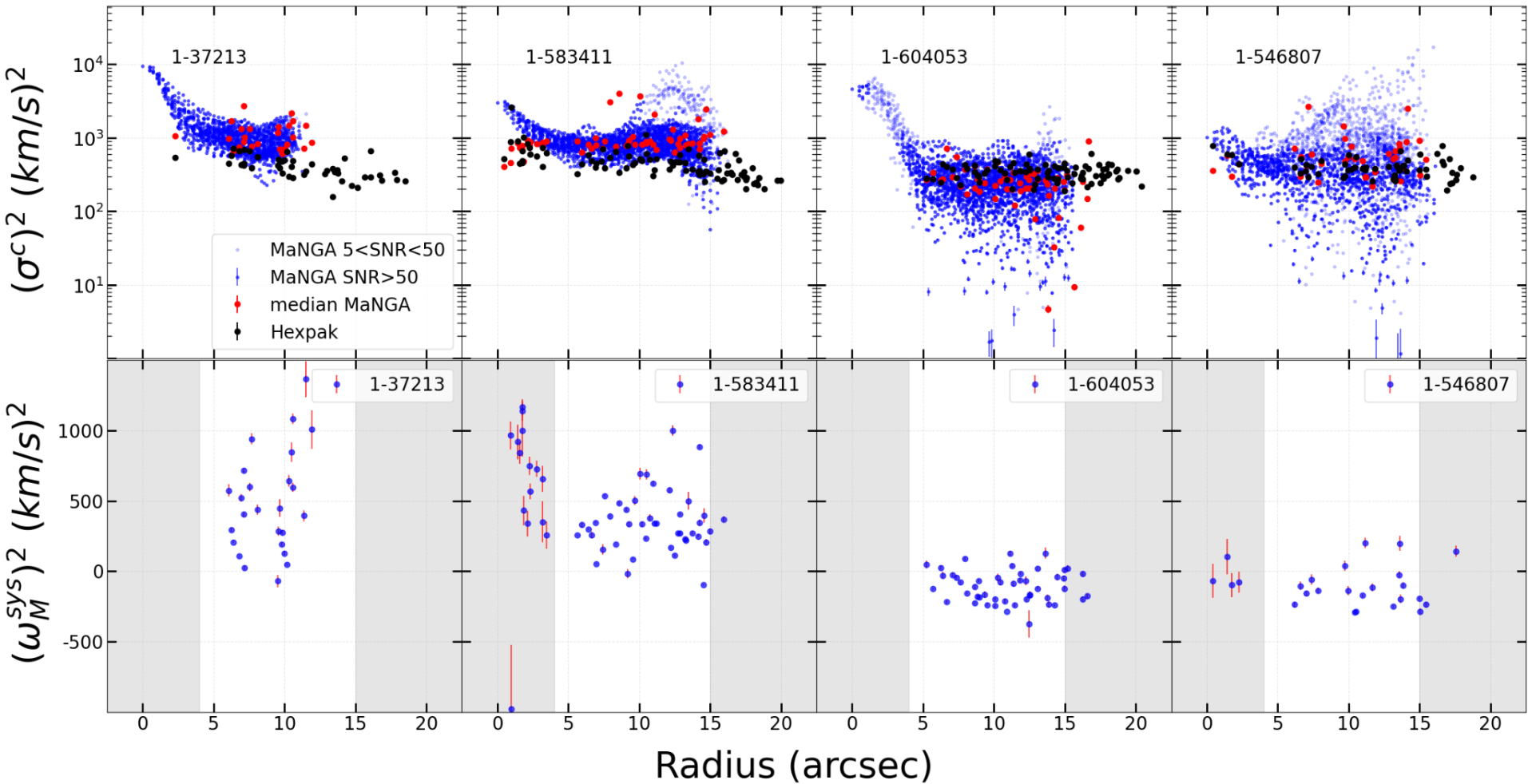}
\caption{Radial distribution of LSF-corrected H$\rm\upalpha$ velocity dispersions (top row) and the square of the systematic error estimates of the MaNGA LSF (bottom row). In the top row individual MaNGA spaxels are shown as blue points, HexPak measurements from individual fibers with SNR$>10$ are shown as black points, and median values for MaNGA spaxels in each of these HexPak fiber footprint where the median MaNGA SNR$>$50 are shown as red points. Systematic error estimates, described in the text, are computed with Equation~\ref{eq:sysigmi_sigm_sigh}. The MaNGA values of $\rm(\upsigma_M^c)^2$ used in this equation are a median of all the spaxels within the footprints of HexPak fibers. Data in the shaded regions (bottom row) are excluded from the median values, $\rm\langle(\upomega_M^{sys})^2\rangle$, for individual galaxies.}
\label{fig:sigma_comp}
\end{figure*}

\begin{figure*}
\centering
\includegraphics[width=\textwidth]{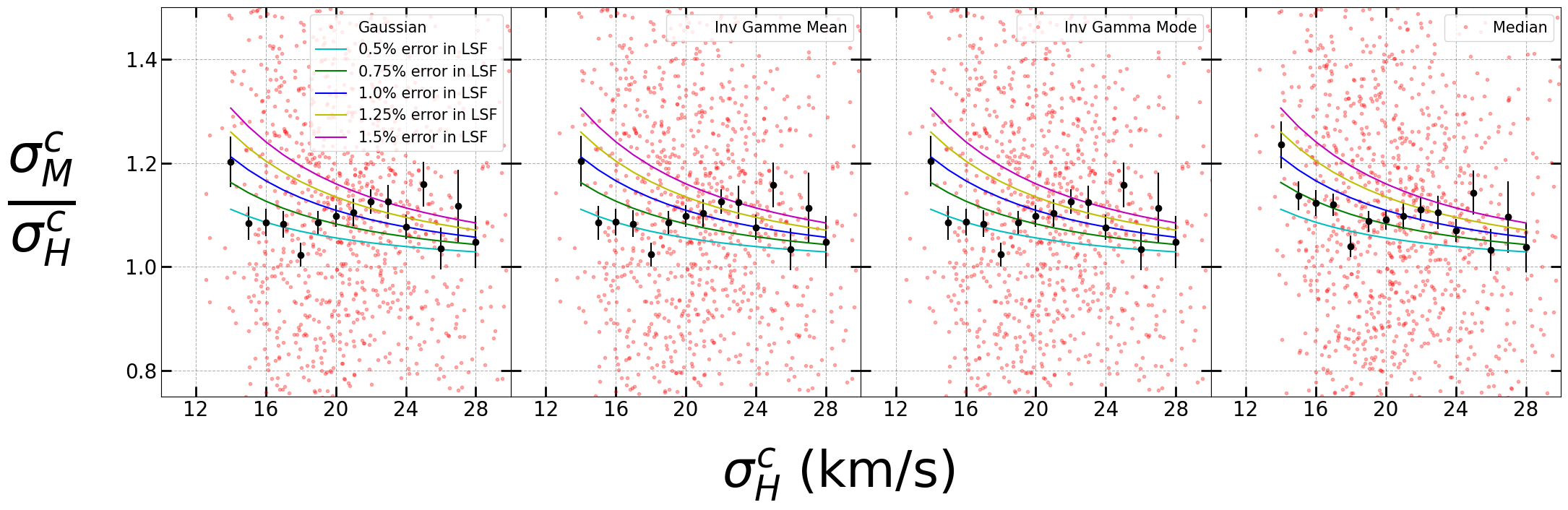}   
\caption{Ratio of LSF-corrected line-widths for MaNGA and HexPak versus the HexPak line-width. Red points represent ratios for individual HexPak fiber footprints with median MaNGA SNR$>$50, as described in the text; the numerator is the median of the aggregate of \textit{all} MaNGA spaxels in the footprint. Black data points represent the mean within a bin of 1~km/s with error bars denoting the error in the mean. Panels (left to right) show ratios computed using bias-corrections adopting mean statistics for different error distributions, and the median. Curves indicate the expected relation with HexPak line-width assuming the MaNGA LSF is systematically \textit{underestimated} by the labelled percentages.}
\label{fig:bias_rem}
\end{figure*}

We examine the results of the simulations in Figure~\ref{fig:bias_sim} using two statistics of the $\rm\upsigma_{\rm est}^2$ distribution. The first, shown in the top row, is the mean of the uncensored values of $\rm\upsigma_{\rm est}$, i.e., the mean of the square-root of all values of $\rm\upsigma_{\rm est}^2>0$. This estimator will contain survival bias; in the case of the IGF error distribution this estimator will also contain positivity bias. The top row is qualitatively similar to what is found in Figure~15 of \cite{Law_2021}, as it should. The second statistic, shown in the bottom row, is the square-root of the median value of $\rm\upsigma_{\rm est}^2$. This is not expected to suffer from survival bias, but in the case of the IGF error distribution, should contain positivity bias.

Comparing the top and bottom rows in Figure~\ref{fig:bias_sim} it is evident that the median statistic has far less systematic bias at any given line-width, SNR, and error distribution function. Indeed, for a Gaussian error distribution, the simulated biases are consistent with zero. With positivity bias, the square median estimation at SNR>=50 has $\rm\leq$ 2\% bias in an astrophysically significant dispersion range, and even at SNR=10 the bias is only $\rm\sim$5\% at $\rm\upsigma_{\rm true}=15$~kms s$^{-1}$ -- a factor of 4.5 below the instrumental resolution.

Simulations such as these can be used to remove biases from line-width measurements, as described for the mean estimator in \cite{Law_2021}. However, bias removal requires good estimates of SNR as well as some ability to guess the intrinsic dispersion, i.e., $\rm\upsigma_{\rm true}$. \cite{Law_2021} suggests that $\rm\upsigma_{\rm true}$ may be estimated by using spaxels within a galaxy that have large SNR, usually 80 or higher. This approach could lead to biases or at least increase in random error in situations where, respectively (i) these spaxels do not well represent the intrinsic dispersion of the lower SNR spaxels being corrected, or (ii) the number of high SNR spaxels is small. The effect of error in the estimated intrinsic dispersion used to assign a correction to the mean and median statistics is shown in Figure~\ref{fig:guess_sim}. Even for an extremely modest error in the estimated intrinsic dispersion (e.g., $\rm\pm1$~km s$^{-1}$), the associated error in the corrected dispersion using the mean statistic becomes quite significant below 20 km s$^{-1}$ even at high SNR in MaNGA data. The corresponding error for the median statistic is much lower simply  because the correction is much smaller. 

These results re-enforce our proposal to use the median statistic based on the squared difference of the observed and instrumental dispersions. Here, and in the development that follows, we work with dispersions as squared values to avoid the issues of `survival bias,' as described in \cite{Law_2021} and in Section~\ref{sec:biases} above; as we have shown in this section working directly with squared values is statistically robust and avoids the complication of modeling error distributions. Figure \ref{fig:bias_sim} demonstrates that by adopting the median statistic of the squared difference of the observed and instrumental dispersions, a SNR threshold of 50 enables recovered dispersions to be within (+3,-2)\% of the astrophysical dispersion at the expected thermal limit of $\sigma=9$ \kms, and roughly 5\% at the same limit for SNR=30. At $\sigma=18$ \kms\ systematics are $<$ 1\% even at SNR=30.

\subsection{Application of the median estimator of line-width to MaNGA and HexPak data}
\label{sec:app}

In detail, at every spaxel, the MaNGA measured dispersion ($\rm\upsigma_M$) is corrected by the DAP-provided pre-pixelized LSF ($\rm\upomega_M$) through quadrature subtraction at the spaxel level:
\begin{equation}
    \rm (\upsigma_M^c)^2 = \upsigma_{M}^2 - \upomega_M^2,
    \label{eq:sigm_sigmo_sigmi}
\end{equation}
assuming, as we do throughout, that the astrophysical and instrumental line-shapes are suitably approximated as Gaussians. Here $\rm\upsigma_M$ is the pre-pixelized Gaussian velocity dispersion, derived from {\tt pPXF} assuming no LSF, and $\rm\upomega_{M}$ is the pre-pixelized LSF Gaussian width -- both as reported by the DAP. In principle we can compute a similar quantity at the fiber level for HexPak data. 

In the present analysis we fit line profiles with a single-component Gaussian model which, in some cases, may be insufficient to parameterize the observed complexity in line-shape. Because we are restricting our analysis to data outside of the steep rise of the rotation curve however, strongly asymmetric line-profiles due to beam-smearing are mostly absent. Further, we find that even for the high-resolution HexPak data the presence of weak, broad-lined components (the topic of a future paper) do not significantly perturb the single component widths. I.e., in two-component fits the narrow-line component is nearly identical in width to that of a single component fit.

Before proceeding to apply the median estimators of $\rm (\upsigma_M^c)^2$ and $\rm (\upsigma_H^c)^2$, we investigated the SNR regime of MaNGA and HexPak data which may introduce survival bias. We did so in the specific context of the analysis here which compares the MaNGA spaxel measurements to HexPak fiber measurements. To make this a fair comparison we compare only spaxels within specific HexPak footprints to the HexPak values, but we include \textit{all} of the MaNGA spaxels within the footprint regardless of the spaxel SNR.\footnote{ An alternative approach would be to coadd the MaNGA spectra and refit the line-width. While this might more closely replicate the HexPak beam-smearing, in practice the MaNGA and HexPak fiber footprints are very similar (1 and 1.5 arcsec fiber radii, respectively), fibers azimuthally scamble the input signal, and the MaNGA spaxels are constructed from a modified Shephard's algorithm that includes contributions from dithered fiber measurements within a radius up to 1.6 arcsec \citep{Law_2016}. Hence the beam smearing in individual MaNGA spaxels measurements are already similar to that of a HexPak fiber. We do not take this approach in order to directly compare with and calibrate extant data from the DAP in the public domain.}

As a practical matter, to proceed with our analysis we will limit our comparison to SNR thresholds in both the single-fiber HexPak data as well as the \textit{median} SNR of MaNGA spaxels within HexPak footprints. In order to understand the implications of this decision on resulting MaNGA SNR distribution we compared the full MaNGA SNR distribution at the spaxel level to the median MaNGA SNR per footprint. Figure \ref{fig:snr_dist} shows these distributions are nearly identical, but that the FWHM of the distribution of SNR values per HexPak footprint is comparable to the median value. This means that by imposing median SNR cuts we will be considering a broad range of MaNGA SNR.

With these SNR statistics in mind, we plot the median of $\rm(\rm\upsigma_M^c)^2$ for all spaxels within each HexPak footprint, denoted $\rm \langle(\upsigma_M^c)^2\rangle$, for HexPak footprints that have SNR>10 within our sample in Figure \ref{fig:sigma_hp_manga}. This figures shows the distribution of the squared values of the HexPak LSF-corrected dispersions at SNR$>$10 is always positive, while the median of the similar quantity for MaNGA spaxels within HexPak footprings do yield significant numbers of negative values. Hence for HexPak data with SNR$>$10 there is no introduction of survival bias from correcting for the LSF broadening. Consequently, for HexPak we henceforth directly compute the LSF-corrected astrophysical dispersion $\rm\upsigma_H^c$ by providing {\tt pPXF} with a template which have sigma equal to the instrumental LSF, $\rm\upomega_H$. However, $\rm\langle(\upsigma_M^c)^2\rangle$ has negative data points at every MaNGA median SNR bin and hence survival bias would be significant had we considered a first-moment estimate (e.g., median) of the linear quantity $\rm\upsigma_M^c$. 

We compare the LSF-corrected H$\rm\upalpha$ emission-line dispersions measured by MaNGA and HexPak in Figure \ref{fig:sigma_comp} for 4 galaxies in our sample. As depicted, the radial trends of LSF-corrected H$\upalpha$ velocity dispersions are qualitatively similar for the two instruments, but for some galaxies the MaNGA measurements can be systematically higher or lower. 

In Figure~\ref{fig:bias_rem} we plot  the ratio of the MaNGA dispersion ($\rm\upsigma^c_M$) to the HexPak dispersions ($\rm\upsigma^c_H$), both LSF corrected, versus the corrected HexPak dispersions for the entire sample. Given the factor $\sim$7 higher spectral resolution we adopt the HexPak measurements as a benchmark. If there were a consistent systematic error in the MaNGA LSF estimates we would expect to see a decreasing trend in the ratio toward larger line-widths, as indicated by the curves. In this Figure we exclude all fibers within r$<4$ arcsec, to avoid beam-smearing issues. Figure~\ref{fig:bias_rem} shows the MaNGA dispersion values computed for three cases using the bias-corrected mean (with the three error distributions described in the previous section) and a fourth case using the median; in all cases the statistics are taken for the set of spaxels within each HexPak fiber footprint, using only footprints where the HexPak SNR$>10$ and the \textit{median} MaNGA SNR$>50$. To estimate the mean bias correction factor in the first three cases, we used the mean LSF-corrected dispersion of MaNGA spaxels with SNR$>$80 for each galaxy. 

While the ratio of LSF-corrected dispersion values for individual fibers has considerable dispersion, the mean values show a clear positive offset consistent with an overall net bias in the MaNGA LSF estimate, which the balance of our analysis will quantify. Further, while the results for the three bias-corrected mean prescriptions are very similar, the median prescription (which does not require a correction) marginally shows the best indication of a trend in the ratio with HexPak dispersion.

\subsection{Systematic errors in the MaNGA instrumental line-width}
\label{sec:sys}

We define the systematic error in $\rm\upomega_{M}$ as $\rm\upomega_M^{sys}$ such that
\begin{equation}
    \rm (\upomega_M^c)^2 = \upomega_M^2 + (\upomega_M^{sys})^2,
    \label{eq:syssigmi_sigmic_sigmi}
\end{equation}
where $\rm\upomega_M^c$ is the systematic-corrected MaNGA LSF, and in principal $\rm(\upomega_M^{sys})^2$ can be either positive or negative. If we assume $\rm\upsigma_H^c$ is much closer to the astrophysical dispersion than $\rm\upsigma_M^c$, we can expect that
\begin{equation}
    \rm (\upsigma_H^c)^2 \sim \upsigma_M^2 - (\upomega_M^c)^2.
    \label{eq:sigh_sigmo_sigmic}
\end{equation}
Consistent with this assumption, Figure \ref{fig:sky_arc_comp} shows that the effect of any variation in measured $\rm\upomega_{H}$ is $\rm\leq$10\% level to $\rm({\upsigma_H^c})^2$. This is due to the fact that $\rm\upomega_{H}$ is expected to be lower than $\rm\upsigma_H^c$. In contrast, this is not the case for MaNGA where $\rm\upomega_{M}$ is larger than $\rm\upsigma_M^c$, and hence even a minor correction in $\rm\upomega_{M}$ leads to large change in $\rm\upsigma_M^c$. 

Combining equations \ref{eq:sigm_sigmo_sigmi}, \ref{eq:syssigmi_sigmic_sigmi} and \ref{eq:sigh_sigmo_sigmic}, we find:
\begin{equation}
    \rm (\upomega_M^{sys})^2 = (\upsigma_M^c)^2 - (\upsigma_H^c)^2.
    \label{eq:sysigmi_sigm_sigh}
\end{equation}
Both $\rm\upsigma_M^c$ and $\rm\upsigma_H^c$ are measured, so we can easily compute $\rm(\upomega_M^{sys})^2$. In practice we use $\rm\langle{(\upsigma}_M)^2\rangle$ within a given HexPak fibre footprint to compute $\rm(\upomega_M^{sys})^2$. Figure \ref{fig:sigma_comp} shows $\rm(\upomega_M^{sys})^2$ against radial distance of the same subset of four MaNGA galaxies discussed earlier. Low signal-to-noise (SNR $\rm\leq 10$) points are excluded in the plot. Measurements in the shaded regions are excluded from our computation of the median systematic corrections; the larger radial distance cut (r $<$ 15\arcsec) is used to ensure spatial overlap with MaNGA data while the lower cut (r $>$ 4\arcsec) ensures spaxels affected with beam smearing are excluded. 

\begin{figure*} 
\includegraphics[width=0.8\textwidth]{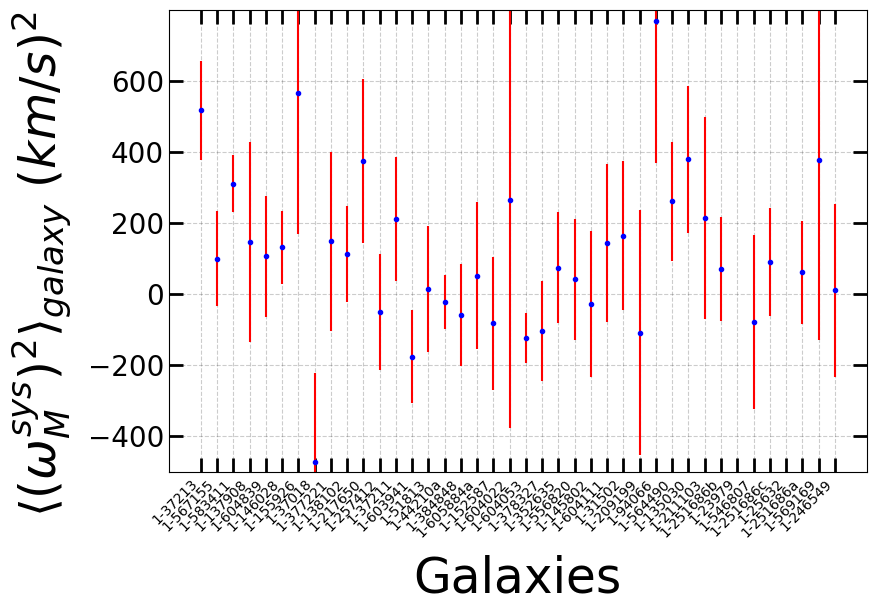} 
\caption{Median of $\rm(\upomega_M^{sys})^2$ for each galaxy in our sample, denoted $\rm\langle(\upomega_M^{sys})^2\rangle_{galaxy}$, computed over all HexPak fibres matched to MaNGA data-cube spaxels that satisfy the conditions that the HexPak SNR$>10$, the median MaNGA spaxel SNR$>$50 for spaxels in the HexPak fiber footprint, and the HexPak fiber radii are between 4 and 15\arcsec. Error bars represent the standard deviation of $\rm(\upomega_M^{sys})^2$.}
\label{fig:syssigma_all}
\end{figure*}

We then compute the median of $\rm(\upomega_M^{sys})^2$ of all the HexPak footprints mapped within each galaxy, denoted as $\rm\langle (\upomega_M^{sys})^2 \rangle_{galaxy}$. Figure \ref{fig:syssigma_all} shows these median values for each galaxy, with distribution histograms given in Figure \ref{fig:histcdf}. The latter also shows the distribution of $\rm(\upomega_M^{sys})^2$ for all of the individual measurements from all galaxies together. For this, we computed $\rm(\upomega_M^{sys})^2$ from Equation~\ref{eq:sysigmi_sigm_sigh} using the median value of $\upsigma_M^c$ for MaNGA spaxels within a given HexPak footprint where the median MaNGA SNR in the footprint is greater than 50 and the HexPak SNR is greater than 10. The HexPak dispersion, $\upsigma_H^c$, is then subtracted in quadrature; we refer to this difference as $\rm(\upomega_M^{sys})^2_{ensemble}$. Although the distribution is somewhat broader for the ensemble, the median values of $\rm\langle (\upomega_M^{sys})^2 \rangle_{galaxy}$ and $\rm(\upomega_M^{sys})^2_{ensemble}$ are identical at 96 $\rm km^2 s^{-2}$, corresponding to a linear value $\rm\langle \upomega_M^{sys} \rangle$ of 9.8~\kms. Additionally the $\rm\langle (\upomega_M^{sys})^2 \rangle_{galaxy}$ distribution has 67\% confidence limit (CL) of $\pm$195 $\rm km^2 s^{-2}$, or 14.0 \kms\ in linear units. This aligns with the error-weighted average and standard deviation of $\rm \langle \upomega_M^{sys} \rangle_{galaxy}$ from figure \ref{fig:syssigma_all} which is 99 and 187 $\rm km^2 s^{-2}$, or 9.9 and 13.7 \kms\ in linear units, respectively. Although in linear units $\rm \langle \upomega_M^{sys} \rangle = 9.8$ \kms\ is a large fraction of the MaNGA LSF, $\rm \upomega_M = 67.6$ \kms, when added in quadrature this amounts to only 1\% increase in MaNGA median LSF estimate. This is a systematic error in the MaNGA estimated LSF in addition to statistical uncertainties which is discussed in the next section.

\section{Implications for astrophysical line-width distributions}
\label{sec:results}

\subsection{Instrumental calibration}
\label{sec:cal}

\begin{figure} 
\includegraphics[width=\columnwidth]{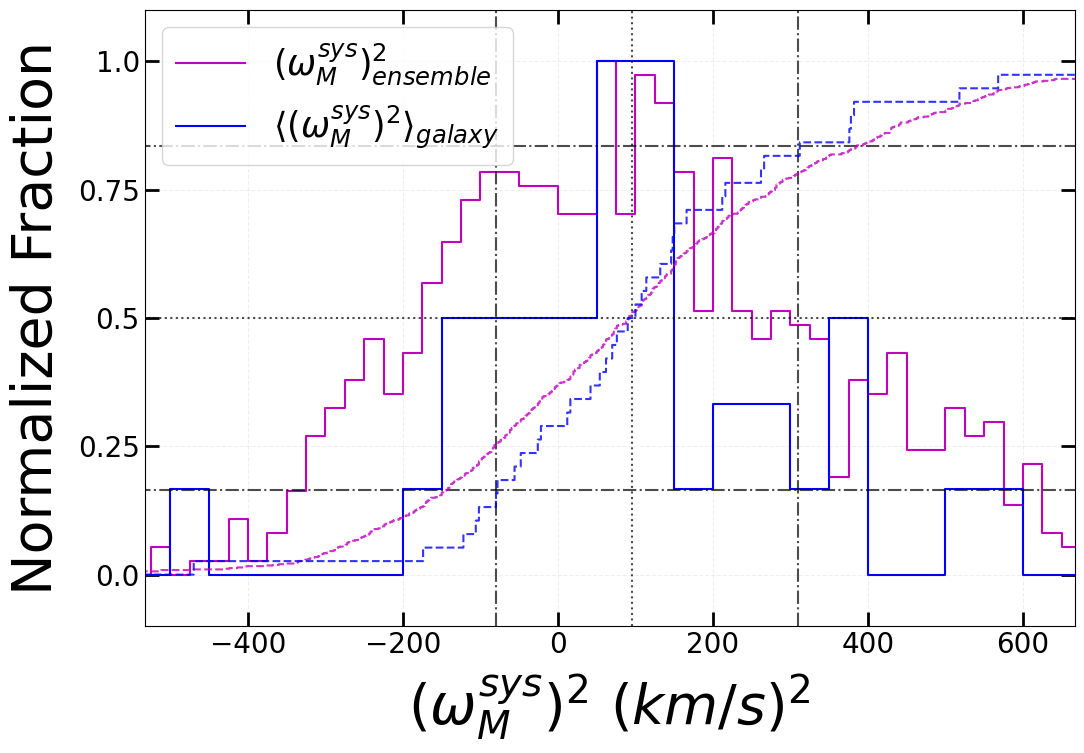} 
\caption{Normalized differential histograms and cumulative distributions (CDF) of $\rm(\upomega_{M}^{sys})^2$ for the median values of each galaxy (blue, $\rm\langle(\upomega_M^{sys})^2\rangle_{galaxy}$), and separately for individual measurements in all galaxies for all HexPak fibers and corresponding MaNGA spaxels that meet SNR and radial criteria given in the text (purple, $\rm(\upomega_M^{sys})^2_{ensemble}$). The 50\% of the CDF for both distributions is marked with a black vertical dotted line at a value corresponding to a median $\rm\upomega_M^{sys}$ of 9.8~\kms ($\rm(\upomega_M^{sys})^2_{galaxy}=96.0$~km$^2$ s$^{-2}$). The black dash dot vertical and horizontal lines denote the 67\% confidence level measured from $\rm(\upomega_M^{sys})^2_{galaxy}$ which is found to be 19.7 \kms.}
\label{fig:histcdf}
\end{figure}

The conclusion from the analysis in the previous section is that the existing MaNGA spectral LSF as reported in \cite{Law_2021} and in DR-17 \citep{DR17} is too small by roughly 1\%, reckoned here at H$\rm\upalpha$ wavelengths: Rather than a median LSF width of 67.6 km s$^{-1}$ at H$\rm\upalpha$ it should be 68.3 km s$^{-1}$. That this external calibration of the MaNGA LSF yields such a small change is a rather remarkable statement about the quality of the DRP \citep{Law_2016, Law_2021}. While a modest correction to the LSF, as we will show below, it does measurably alter the line-width distribution for HII-like regions within MaNGA galaxies, particularly demonstrating a 25\% decrease in kinetic energy. 

Further, from Figures~\ref{fig:syssigma_all} and \ref{fig:histcdf}, it appears that there is real variation between MaNGA data-cubes, with a 67\% confidence level of $\pm$14~\kms\ about $\rm\langle \upomega_M^{sys} \rangle
_{galaxy}$. This implies that while the calibration here should serve to accurately estimate astrophysical velocity dispersions from MaNGA data in the mean, the measured distribution of these widths will be  broadened by roughly 14~\kms\ over the underlying astrophysical distribution width.

\begin{figure} 
\includegraphics[width=\columnwidth]{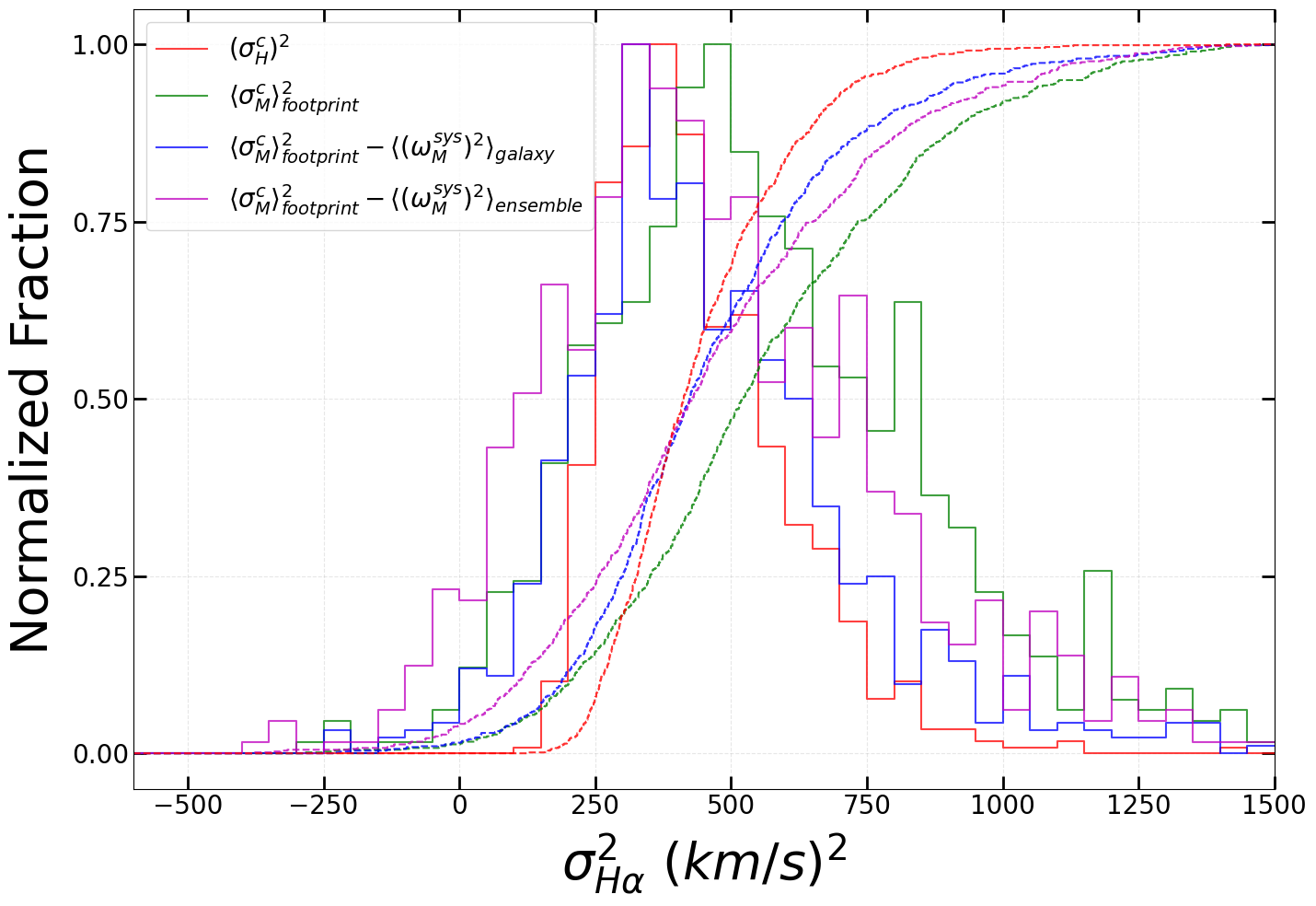} 
\caption{Normalized differential histograms and CDFs comparing the distribution of the LSF-corrected $\rm\upsigma_{H\upalpha}$ for HexPak and MaNGA data, with and without corrections for systematic errors in the MaNGA LSF. All MaNGA spaxels located within a HexPak footprint (with median MaNGA SNR $>$50) at radial distance between 4\arcsec\ and 15\arcsec\ are used. Red curves represent $\rm(\upsigma_H^c)^2$: the distribution of all LSF-corrected Hexpak fiber measurements with SNR$>$10. Green curves represent $\rm\langle(\upsigma_M^c)^2\rangle$: the distribution of LSF-corrected MaNGA spaxel measurements without any systematic correction to the LSF. Magenta curves represent $\rm\langle(\upsigma_M^c)^2\rangle - \langle(\upomega_M^{sys})^2\rangle_{\rm ensemble}$: the distribution of of MaNGA spaxel measurements with a systematic correction to the MaNGA LSF using the median value of $\rm\langle(\upomega_M^{sys})^2\rangle$ for our sample. Blue curve represents $\rm\langle(\upsigma_M^c)^2\rangle - \langle(\upomega_M^{sys})^2\rangle_{\rm galaxy}$: the distribution of of MaNGA spaxel measurements with systematic correction to the MaNGA LSF of respective galaxy using the $\rm\langle(\upomega_M^{sys})^2\rangle$ of that galaxy.}
\label{fig:sysigmacorr_hist}
\end{figure}

To illustrate the impact of the variations in the MaNGA LSF systematic, Figures~\ref{fig:sysigmacorr_hist} displays the distribution of corrected H$\upalpha$ line-widths computed in three different ways. We continue to use only HexPak fibers with SNR>10 and MaNGA spaxels within the HexPak footprints with median MaNGA SNR>50 to ensure the comparison is consistent with earlier analysis. As before, the HexPak values ($\rm(\upsigma_H^c)^2$, in red) serve as the benchmark both for the distribution median and width.

(i) First we compare this distribution to the LSF-corrected MaNGA line-width defined by Equation~\ref{eq:sigm_sigmo_sigmi} ($\rm (\upsigma_M^c)^2$, in green) using the nominal LSF values from the DAP, \rm $\rm\upomega_M$; the distribution clearly has both a larger median and width. 

We then recompute the LSF-corrected MaNGA line-width to take into account our estimated systematic correction to the MaNGA LSF by substituting $\rm\upomega_M^c$ for $\rm\upomega_M$ in Equation~\ref{eq:sigm_sigmo_sigmi}. Following Equation~\ref{eq:syssigmi_sigmic_sigmi} we can do this in two ways by assigning $(\upomega_M^{sys})^2$ either (ii) to the median value from Figure~\ref{fig:histcdf}, i.e., the same correction for every measurement (purple); or (iii) to $\rm\langle (\upomega_M^{sys})^2 \rangle_{galaxy}$ for measurements on a galaxy by galaxy basis (blue). 

Both of the distributions corrected for the LSF systematic (cases ii and iii) have median values that closely match the HexPak value, which follows from the results of Figure~\ref{fig:histcdf}. However, when applying the systematic correction to the MaNGA LSF on a galaxy by galaxy basis (case iii), the width of the distribution narrows and comes into closer agreement with the HexPak distribution width. Indeed, the 
standard deviation derived from the 67\% confidence range, in linear units, decreases from 16.9~\kms\ to 15.7~\kms (cases ii to case iii), compared to 12.4~\kms for HexPak. The difference in the distribution width between these two cases is comparable to the differences seen in Figure~\ref{fig:histcdf}, as would be expected. That the case (iii) distribution remains substantially broader than the HexPak distribution might suggest there remain uncorrected LSF systematic variations with spatial position within each data-cube; as seen in Figure 12 of \cite{Law_2021}, LSF spatial variations are present due to the mapping of fibers to different spectrograph slit blocks. With the limited re-calibration offered by this study, the broader distribution represented by the purple curve is what can be expected for the full MaNGA sample SNR$>$50.

% From Sabya. For Fig 15
%                      (0.165-0.5) and (0.5-0.835)
%MaNGA using omega_M (green curve) -- 249 and 318
%MaNGA using (omega_M^c)_ensemble (purple) -- 250 and 318 <>=284
%MaNGA using (omega_M^c)_galaxy (blue) -- 180  and 252 <>=216
%HexPak (red curve) -- 124 and 186 <>=155

\begin{figure} 
\includegraphics[width=\columnwidth]{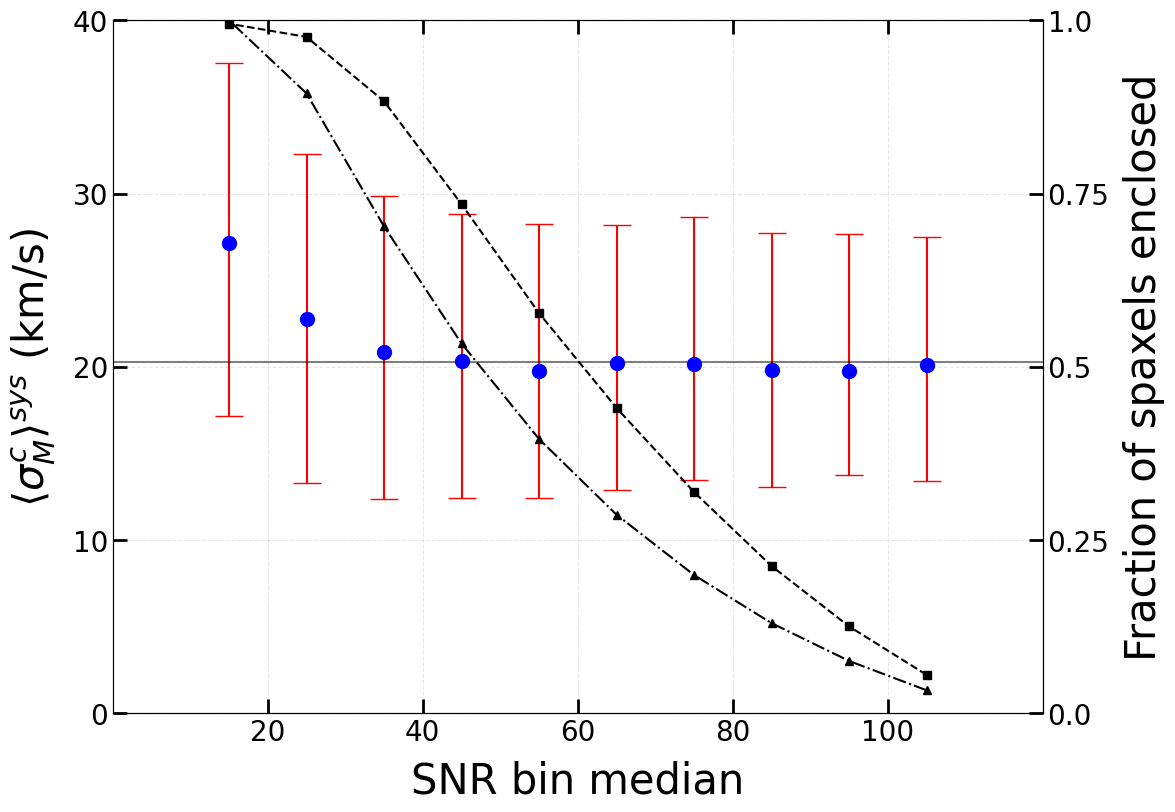} 
\caption{Median and 67\% confidence range of the fully-corrected MaNGA H$\upalpha$ line-widths of all star-froming MaNGA spaxels, $\rm\langle\upsigma_M^c\rangle^{sys}$ (Equation~\ref{eq:syscor_sigma}), located between 4\arcsec-15\arcsec\ radially from the center, versus spaxel SNR. The solid black line shows the HexPak median for HexPak fibers with SNR>10. The trend (or lack thereof) in the median points for MaNGA data shows that the SNR threshold of MaNGA spaxels for reliable line-width measurements can be as low as SNR=30. The dashed and dash-dot lines represents cumulative number of spaxels included as a function of decreasing SNR for our sample and for all MaNGA star-forming spaxels respectively. A SNR\>50 cutoff includes only 57\% of spaxels in our sample while SNR>30 includes almost 88\% of the spaxels. The number of included spaxels are slightly lower at 39\% and 70\% for the entire MaNGA sample at SNR cutoffs of 50 and 30 respectively.}
\label{fig:sysigmacorr_snr}
\end{figure}

\subsection{Limiting SNR}
\label{sec:limsnr}

The results from simulations in Figure~\ref{fig:bias_sim} indicate that the median estimator may remain an accurate measure of line-width below the recommended cutoff of SNR=50 when using the corrected mean formulation from \cite{Law_2021}. To test this we can make the astrophysical assumption that if we examine the distribution of line-widths from MaNGA spaxels with HII-like line-ratios, the median width should be independent of SNR. However, the spaxel SNR in the H$\upalpha$ line for MaNGA data correlates with the star-formation surface-density ($\rm\upSigma_{SFR}$). As \cite{Law_2022} and many others have shown, and as we will explore in the following section, there is indeed a correlation between line-width and $\rm\upSigma_{SFR}$, whereby line-width increases with $\rm\upSigma_{SFR}$. Nonetheless the trend is sufficiently shallow that over a modest range in SNR our assumption of near-constancy in line-width for HII-like spaxels should suffice.

\begin{figure*} 
\includegraphics[width=\textwidth]{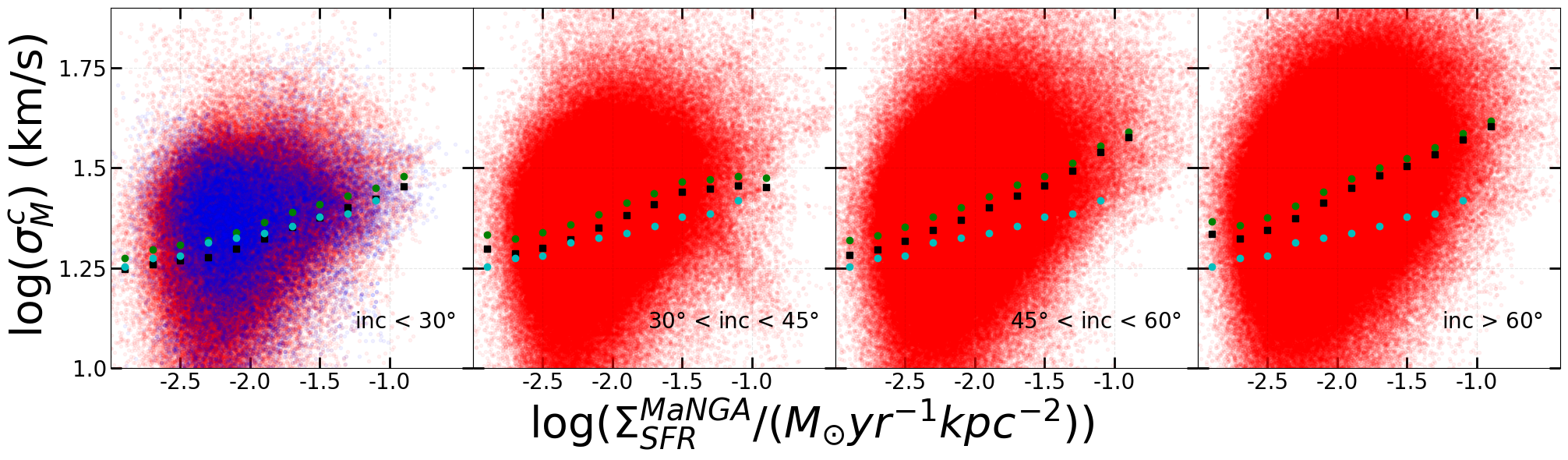} 
\caption{Effect of LSF correction on dispersion - star formation rate surface brightness relation. The red points indicate all MaNGA spaxels after implementing the selection criteria described in the text. The four panels separates the inclination effect. Black and green  points are binned mean of all MaNGA spaxels with and without systematic corrections to the MaNGA LSF. The overlaid blue points are MaNGA spaxels of the galaxies in our HexPak sample. The cyan points are the binned mean relation obtained from HexPak observations of our sample of face-on (inc < 30\degree) galaxies, repeated in all panels. HexPak and MaNGA binned mean shows impressive agreement in low inclination (left panel, inc < 30\degree) and slowly drifts away as the inclination increases demonstrating the effect of inclination in $\upsigma-\upSigma$ relation.}
\label{fig:SFRcomp}
\end{figure*}

We select MaNGA spaxels in the same way as \cite{Law_2022} using these criteria:
\begin{enumerate}
%    \item $\rm SNR > 50$ for H$\rm\upalpha$,
    \item $\rm SNR > 3$ for H$\rm\upbeta$, [OIII], [NII], and [SII],
    \item $\rm \rm\upsigma_{H\upalpha}$ < 100 km s$^{-1}$,
    \item radii $>4$ arcsec to avoid beam smearing effects, and
    \item line-emission is consistent with star-forming regions defined by \cite{bpt} based on line-ratios (equation 2 from \cite{Law_2021b}).
\end{enumerate}
Rather than limiting SNR$>50$ as done by \cite{Law_2022}, we aggregate
measurements of line-width, made at the spaxel level, in bins of SNR. Here we use all spaxels within 4\arcsec-15\arcsec radius for all galaxies in our sample, regardless of whether they fall within a HexPak footprint. We then compute the median and 67\% CL of the  LSF-corrected line width in each bin, using our systematic correction to the MaNGA LSF:
\begin{equation}
\rm\langle\upsigma_M^c\rangle^{sys} \equiv \sqrt{\langle(\upsigma_M^c)^2 - (\upomega_M^{sys})_{\rm ensemble}^2\rangle}.
\label{eq:syscor_sigma}
\end{equation}
We specifically use $\rm(\upomega_M^{sys})_{\rm ensemble}$ to understand the behavior of the corrected line-width as it would be applied to any of the MaNGA data.

Figure \ref{fig:sysigmacorr_snr} shows the change with SNR in the median values of $\rm\langle\upsigma_M^c\rangle^{sys}$ and the uncertainty corresponding to the 67\% CL. The flat profile with SNR shows that the median estimator applied to the squared difference of observed line-width and LSF is constant over a factor of $\sim$3 in SNR, down to a SNR threshold $>$30. The upturn at SNR$<$30 may reflect an astrophysical effect at low $\rm\upSigma_{SFR}$, or contamination from a broader-lined, diffuse ionized gas (DIG) component at low SNR where the culling based on line ratios is more uncertain. By reducing the SNR threshold from 50 to 30, this increases the fraction of all MaNGA spaxels for gas line-width measurements by nearly a factor of two from 39\% to 70\%.

\subsection{Effect on astrophysical measurement: the star-formation rate - dispersion correlation}
\label{subsec:astro}

As previously noted, \cite{Law_2022} and references therein have demonstrated that the velocity dispersion of ionized gas and star-formation rate surface-density ($\rm\upSigma_{SFR}$) -- both using H$\rm\upalpha$ -- are correlated, albeit with significant scatter. This scatter is likely due to a combination of measurement error, variations in the coupling efficiency of the radiative and mechanical energy from star formation to the gas, and also geometric effects. We are careful to consider the effects of inclination on our line-of-sight measurements of $\rm\upSigma_{SFR}$ and $\rm\upsigma_{H\upalpha}$ since our sample is comprised of mostly face-on galaxies while the MaNGA sample, and galaxy samples overall often have, a wide range of inclinations.

By virtue of the enormous MaNGA sample of galaxies selected at all inclination and the two dimensional spectral coverage, \cite{Law_2022} were able to disentangle some of the geometric effects. In their study a beam-smearing correction is applied to all spaxel measurements to account for the line-of-sight integration through the projected velocity field of an inclined, rotating disk.  After making this correction, \cite{Law_2022} find evidence for an anisotropic velocity ellipsoid for the ionized gas with the vertical component roughly 10 to 15\% lower than in the in-line components; i.e., the velocity ellipsoid is slightly flattened (and somewhat triaxial). 

In our analysis we have not implemented a beam-smearing correction. Both beam-smearing and a flattened velocity ellipsoid will tend to make the projected velocity dispersion larger at higher inclination, while line-of-sight integration will tend to make the projected surface-brightness ($\rm\upSigma$) larger by $1/\cos(i)$. These two effects tend to compensate in terms of the $\rm\upSigma-\upsigma$ correlation, but also distend the spaxel distribution in this space with increasing inclination.

To illustrate these effects due to inclination, we selected MaNGA spaxels in the same way as  given in the previous section with the addition here of $\rm SNR > 50$ for H$\rm\upalpha$  to be fully consistent with \cite{Law_2022}. We also used the prescription to compute the $\rm\upSigma_{SFR}$ described Equation 2 in \cite{Law_2022}. We used H$\rm\upalpha$ flux and angular distance measured by MaNGA DAP to compute the H$\rm\upalpha$ luminosity. We also computed the scale factor representing the solid angled posed by 0.5 arcsec wide MaNGA spaxels. Since we are calculating a surface-density, the quantity is independent of the expansion rate, H$_0$. The  H${\upalpha}$ flux provided by the DAP is already corrected for Milky Way extinction, but not for internal extinction from the host galaxy. Hence we corrected the H${\upalpha}$ luminosity for the latter using the prescription provided by \cite{Cardelli} and the Balmer decrement assuming an intrinsic value of H$\rm\upalpha$/H$\rm\upbeta$ = 2.86. We then applied the systematic correction in $\rm\upsigma^c_M$ to understand the effect of the systematic correction to the MaNGA LSF on the $\rm\upsigma$ vs $\rm\upSigma_{SFR}$ relation. 

Figure \ref{fig:SFRcomp} illustrates the effect of our systematic correction to the MaNGA LSF for $\rm\upsigma^c_M$ in galaxies at different inclinations to our line of sight. The red data points denotes LSF corrected dispersion in linear space (i.e. with survival bias) of all MaNGA spaxels within the above itemized criteria, but for different ranges of inclination in each panel. The blue points are MaNGA spaxels from our sample following the same criteria, repeated in all four panels for comparison; these galaxies are predominantly at low inclination (b/a $>$ 0.7) as evident by the good match to the full MaNGA sample for $i<30^\circ$. 

Comparison of HexPak and  MaNGA average dispersions in bins of star-formation surface-density in Figure \ref{fig:SFRcomp} emphasize the effect of inclination and the systematic correction to the MaNGA LSF: Green circles are $\rm\overline{\upsigma^c_M}$, the 2.5$\rm\upsigma$ clipped average of all MaNGA data in a given inclination bin. Black squares are $\rm\overline{(\upsigma^c_M)^{sys}}$, the same statistics applied after the data are corrected at the spaxel level by $\rm(\langle\upomega^{sys}_M\rangle)^2$. The cyan circles denotes the binned average of HexPak observations, also LSF corrected, which are repeated in each panel for reference. We computed the HexPak star formation surface brightness ($\rm \upSigma^{HexPak}_{SFR}$) via calibrating the HexPak flux against the median $\rm \upSigma^{MaNGA}_{SFR}$ within individual HexPak footprint using a linear relation. We then use this relation to convert the binned $\rm \upSigma^{MaNGA}_{SFR}$ to $\rm \upSigma^{HexPak}_{SFR}$ to locate the cyan points representing the median of $\rm \upsigma^c_H$ within the same surface brightness interval.

Qualitatively, as expected (1) with larger inclination, the binned average of the MaNGA sample deviates from the HexPak (low-inclination) sample and becomes steeper; (2) the correction factor is more dominant in the lower $\rm\upsigma_M^c$--lower $\rm\upSigma_{SFR}$ region, making the relation steeper particularly in the higher inclination bins which is due to the sample selection and not a bias in the MaNGA lSF; and (3) the face-on HexPak sample produces an identical relation to that MaNGA spaxels in galaxies with inclination <30$\rm\degree$.

To quantify these effects, we fit a linear relation between $\rm\log\upsigma_{H\upalpha}$ and $\rm\log\upSigma_{SFR}^{MaNGA}$ for each of $\rm\upsigma_{H\upalpha} = \overline{\upsigma^c_M}$, $\rm \overline{(\upsigma^c_M)^{sys}}$, and $\rm\upsigma_H^c$ over the range $-0.2<\rm\log\upSigma_{SFR}^{MaNGA}<1.2$ that is well sampled by the MaNGA data:
\begin{equation}
\begin{aligned}
    \rm \log \upsigma_{H\upalpha} = \log\upsigma_{-2.5} + p\times\left[2.5+\log\left(\upSigma_{SFR}^{MaNGA}\right)\right]
\end{aligned}
\end{equation}
where p describes the power-law index between 
$\rm\upsigma_{H\upalpha}$ and $\rm\upSigma_{SFR}^{MaNGA}$, and $\rm\log\upsigma_{-2.5}$ (the intercept) is the fitted value of $\rm\log\upsigma_{H\upalpha}$ at $\rm\upSigma_{SFR}^{MaNGA}=-2.5$. We fit all MaNGA spaxels that follow the above criteria in separate inclination bins. The results are found in Table \ref{tab:powerlawfit}. The fit parameters at low inclination for all measures agree well, with systematic increase in both p and $\rm\log\upsigma_{-2.5}$ with inclination. For $i<30^\circ$, p is identical for $\rm\overline{\upsigma^c_M}$ and $\rm\overline{(\upsigma^c_M)^{sys}}$, and slightly steeper than for $\rm\upsigma_H^c$, but the zeropoint $\rm\upsigma_{-2.5}$ for $\rm\overline{(\upsigma^c_M)^{sys}}$ matches to $\rm\upsigma^c_H$ within 1-$\upsigma$ error.

\begin{table}
%\scriptsize
\centering
\caption{Power-law model parameters for gas line-width $\rm\upsigma$ and star-formation surface-density $\rm\upSigma_{SFR}$}
\begin{tabular}{@{}cccc@{}}
\toprule
& \textbf{Inclination} & \textbf{p} & \textbf{$\rm\upsigma_{-2.5}$ (\kms)} \\
\midrule
\multirow{4}{*}{$\rm\overline{\upsigma_M^c}$}& 0\degree - 30\degree & 0.11 $\pm$ 0.07 & 20.0 $\pm$ 0.95  \\
& 30\degree - 45\degree & 0.13 $\pm$ 0.04 & 20.9 $\pm$ 0.57  \\
& 45\degree - 60\degree & 0.13 $\pm$ 0.03 & 22.4 $\pm$ 0.42 \\
& 60\degree - 90\degree & 0.16 $\pm$ 0.08 & 23.4 $\pm$ 1.27  \\
\midrule
\multirow{4}{*}{$\rm\overline{(\upsigma_M^c)^{sys}}$} & 0 \degree - 30\degree & 0.11 $\pm$ 0.1 & 17.8 $\pm$ 1.27 \\
& 30\degree - 45\degree & 0.14 $\pm$ 0.05 & 20.0 $\pm$ 0.65 \\
& 45\degree - 60\degree & 0.14 $\pm$ 0.03 & 20.9 $\pm$ 0.46  \\
& 60\degree - 90\degree & 0.17 $\pm$ 0.09 & 22.4 $\pm$ 1.52  \\
\midrule
$\rm\upsigma_H^c$   & & 0.09 $\pm$ 0.06  & 18.6 $\pm$ 0.81 \\
\bottomrule
\end{tabular}
\label{tab:powerlawfit}
\end{table}

\section{Conclusions}
\label{sec:conclusions}

We have measured the systematic error in previous estimates of the MaNGA LSF in the H$\upalpha$ wavelength region by directly comparing MaNGA measurements of ionized gas line-widths to independent measurements from high-resolution data taken with the HexPak IFU. Uncertainties in the HexPak instrumental resolution are sufficiently small to allow us to calibrate the MaNGA instrumental resolution to better than 0.1\%.

We have also developed an approach to correcting observed line-widths for instrumental broadening by working with the squared differences of observed and instrumental dispersions rather than their square root. This avoids issues with survival bias as well as associated uncertainties for its correction due to modeling the random error distribution. Based on simulations we find that the square root of the median value of the corrected line-width (i.e., squared differences) is an unbiased measure of the median of the astrophysical line-width with $<$2\% systematic error down to SN=30 at astrohphysically relevant dispersions (a thermal broadening limit of $\upsigma=9$~\kms) regardless of the detailed random error distribution. We refer to this as the squared-median method.

The squared-median method has been applied to a sample of low inclination, star-forming galaxies in a radial region where beam-smearing effects are negligible. For MaNGA spaxels spatially co-located within the larger HexPak fiber footprints where both the HexPak fiber SNR$>$10 and the median MaNGA spaxel SNR$>$50, we find the MANGA LSF is underestimated by $\sim$1\% at H$\rm\upalpha$ wavelengths \textit{on average}. This is equivalent to a $\rm\sim$9.8~\kms\ correction added in quadrature to the nominal MaNGA LSF of 67.6~\kms, yielding a corrected LSF of 68.3~\kms\ at H$\upalpha$ wavelengths. This correction is remarkably small, which is testament to the careful calibration in the MaNGA DRP \citep{Law_2016, Law_2021}.

There remains a real variation in the LSF systematic that varies from galaxy to galaxy and likely within the different spatial elements sampling a single galaxy. This is not surprising given the discrete changes in the MaNGA LSF between integral-field fiber bundles and within fiber bundles that map to different discrete slit-blocks in the BOSS spectrographs \citep{Law_2021}. The re-calibration of the MaNGA LSF here, since it is based on a handful of galaxies, cannot address these variations for the full MaNGA sample. While the systematic correction to the LSF above should yield accurate median corrected astrophysical line-widths, we estimate that any distribution function of astrophysical line-widths is likely broadened by a 14~\kms\ dispersion due to these galaxy-to-galaxy and internal LSF variations.

Application of this systematic correction to the MaNGA instrumental dispersion yields two pertinent results. First, the median corrected (astrophysical) dispersion for HII-like line-emission in our calibrator sample, binned by SNR in the H$\upalpha$ line at the spaxel level, is found to be constant at 20~\kms\ between $\rm 30<SNR<110$. This suggests that the squared-median method can be  applied robustly to a limiting SNR of 30, yielding nearly a factor of two increase in spaxels available for kinematic measurements compared to earlier MaNGA studies limited to SNR$>$50. Second, we have revisited the correlation between ionized gas velocity dispersion and star-formation surface-density ($\upSigma_{SFR}$). Here, we have analyzed the full MaNGA data, limited by SNR$>$50 for direct comparison with \cite{Law_2022}. As expected, there is an inclination dependence to the relation, primarily due to line-of-sight effects on the observed line-width. For the subset of MaNGA galaxies with 0\degree - 30\degree inclination (comparable to our calibrator sample), we find power-law fits that are comparable for HexPak and MaNGA data. The agreement becomes marginally better when the systematic correction to the MaNGA LSF is applied. This indicates the results from \cite{Law_2022} are robust to these small changes in the MaNGA LSF.

Given the small variations in recovered (LSF-corrected) line widths for nebular emission over a broad range in wavelength in MaNGA data \citep{Law_2021}, the mean correction determined here can likely be applied across all wavelengths as a zeropoint shift in the MaNGA LSF vector. Future analysis of high-resolution line-widths measurements for H$\upbeta$ and $\rm [OIII]$ will verify this and extend the calibration to stellar velocity dispersions.

\section*{Acknowledgements}
MAB acknowledges support from NSF AST-1517006 and AST-1814682.
SC and MAB acknowledge support from NRF/SARChI-114555.

Funding for the Sloan Digital Sky Survey IV has been provided by the Alfred P. Sloan Foundation, the U.S. Department of Energy Office of Science, and the Participating Institutions. SDSS-IV acknowledges support and resources from the Center for High-Performance Computing at the University of Utah. The SDSS website is www.sdss.org.

SDSS-IV is managed by the Astrophysical Research Consortium for the Participating Institutions of the SDSS Collaboration including the Brazilian Participation Group, the Carnegie Institution for Science, Carnegie Mellon University, the Chilean Participation Group, the French Participation Group, Harvard-Smithsonian Center for Astrophysics, Instituto de Astrofísica de Canarias, The Johns Hopkins University, Kavli Institute for the Physics and Mathematics of the Universe (IPMU)/University of Tokyo, the Korean Participation Group, Lawrence Berkeley National Laboratory, Leibniz Institut für Astrophysik Potsdam (AIP), Max-Planck-Institut für Astro- nomie (MPIA Heidelberg), Max-Planck-Institut für Astrophysik (MPA Garching), Max-Planck-Institut für Extraterrestrische Physik (MPE), National Astronomical Observatories of China, New Mexico State University, New York University, Uni- versity of Notre Dame, Observatário Nacional/MCTI, The Ohio State University, Pennsylvania State University, Shanghai Astronomical Observatory, United Kingdom Participation Group, Universidad Nacional Autónoma de México, Univer- sity of Arizona, University of Colorado Boulder, University of Oxford, University of Portsmouth, University of Utah, University of Virginia, University of Washington, University of Wisconsin, Vanderbilt University, and Yale University.

\section*{Data Availability}
The data underlying this article will be shared on reasonable request to the corresponding author.



\bibliographystyle{mnras}
\bibliography{ref} 

%%%%%%%%%%%%%%%%%%%%%%%%%%%%%%%%%%%%%%%%%%%%%%%%%%

%%%%%%%%%%%%%%%%% APPENDICES %%%%%%%%%%%%%%%%%%%%%

\appendix

\section{Target Sample}
\label{app:targets}

Table~\ref{tab:sample} lists the MaNGA galaxies observed in this HexPak program. The table containts the MaNGA ID (MID), plate and IFU from the MaNGA survey, celestial coordinates, some salient photometric properties, the spectroscopic redshift, and the HexPak run. The Sersic index (n$_S$), total stellar mass (M*) and absolute magnitudes in $g$ and $i$ bands (adopting $\rm\upomega_m=0.3$, $\rm\upomega_\Lambda=0.7$, $H_0=100$ km s$^{-1}$ Mpc$^{-1}$), rest-frame NUV-r color, the half-light radius ($r_{05}$) and redshift are taken from NASA-Sloan Atlas \citep[NSA;][]{Blanton_2011}. Magnitudes are given in the \textit{AB} system.

\begin{table*}
\caption{Galaxy Sample}
\label{tab:sample}
%\tablecaption{Galaxy Sample}
%  \tablehead{
\begin{tabular}{llcccccccccc}
\hline
MID & plate-IFU & RA (2000) & DEC (2000) & n$_S$ & M*                         &${\rm M_i}-5\log h$&${\rm M_g}-5\log h$&NUV-r&r$_{05}$ & z & Run$^c$ \\
    &           & (h:m:s)   & (d:m:s)    &       & ($10^{9} {\rm M}_\odot h^{-1}$)  &  (mag)    &  (mag)    &(mag)&(arcsec) &   & \\
 \hline
1-31502$^a$  &   8656-1901    &   00:30:52.176   &   00:31:43.53   &   1.330   &   1.289   &   -18.351   &   -17.816   &   1.916   &   6.372   &      0.019   &   C   \\
1-37018      &   8077-12704   &   02:45:07.336   &   00:57:00.59   &   2.678   &   7.474   &   -19.998   &   -19.229   &   2.885   &   9.812   &      0.025   &   A   \\
1-37213      &   8078-6103    &   02:49:39.970   &  -00:04:11.46   &   1.902   &   23.87   &   -21.172   &   -20.355   &   3.057   &   6.417   &      0.028   &   A   \\ 
1-37211      &   8078-12703   &   02:50:16.864   &   00:05:31.16   &   2.062   &   29.76   &   -21.335   &   -20.480   &   2.874   &   10.696  &      0.028   &   A   \\
1-603941$^a$ &   8154-9102    &   03:03:50.459   &  -00:12:16.04   &   0.953   &   12.87   &   -20.782   &   -20.173   &   2.306   &   8.175   &      0.028   &   C   \\
1-604022$^a$ &   8080-12703   &   03:17:57.072   &  -00:10:08.76   &   1.111   &   32.46   &   -21.543   &   -20.794   &   2.468   &   11.072  &      0.023   &   C   \\
1-583411     &   8083-12702   &   03:20:58.900   &  -00:22:03.66   &   2.642   &   29.78   &   -21.730   &   -21.023   &   2.578   &   28.865  &      0.021   &   A   \\ 
1-604053     &   8083-12704   &   03:22:47.228   &   00:08:57.72   &   0.986   &   17.66   &   -20.982   &   -20.251   &   2.282   &   14.472  &      0.023   &   A,C  \\
1-38618$^b$  &   8084-1902    &   03:30:29.420   &  -00:29:19.57   &   2.698   &   2.270   &   -18.446   &   -17.402   &   4.907   &   2.231   &      0.022   &   A   \\
1-604111     &   9190-12703   &   03:37:58.867   &  -06:16:14.30   &   1.139   &   22.59   &   -21.035   &   -20.151   &   2.129   &   15.797  &      0.022   &   C   \\ 
1-51813      &   8727-12702   &   03:39:34.894   &  -06:02:19.93   &   0.682   &   6.836   &   -20.078   &   -19.429   &   2.639   &   12.835  &      0.022   &   C   \\
1-377221     &   8132-12702   &   07:23:33.243   &   41:26:05.66   &   3.026   &   36.51   &   -21.603   &   -20.771   &   2.457   &   17.813  &      0.028   &   A   \\
1-378327     &   8239-6101    &   07:41:12.977   &   47:40:17.32   &   1.270   &   5.932   &   -19.786   &   -19.113   &   2.619   &   14.168  &      0.021   &   C   \\
1-152587$^a$ &   8145-12704   &   07:44:57.444   &   28:55:39.01   &   0.821   &   4.365   &   -19.772   &   -19.109   &   2.125   &   8.844   &      0.023   &   C   \\
1-217650     &   8726-3703    &   07:47:41.147   &   22:46:53.78   &   1.450   &   11.75   &   -20.423   &   -19.632   &   2.778   &   4.993   &      0.028   &   A   \\ 
1-604839     &   8146-12701   &   07:49:05.806   &   28:37:09.77   &   0.676   &   13.97   &   -20.847   &   -19.997   &   2.396   &   12.353  &      0.028   &   A   \\
1-145802     &   8148-9102    &   07:51:42.169   &   27:36:26.62   &   1.794   &   2.691   &   -19.563   &   -19.021   &   2.131   &   7.874   &      0.026   &   C   \\
1-379255$^b$ &   8711-1901    &   07:53:03.975   &   52:44:35.53   &   4.694   &   1.482   &   -17.915   &   -16.882   &   5.122   &   1.358   &      0.018   &   A   \\
1-146028     &   8147-12703   &   07:54:22.231   &   27:00:31.70   &   0.684   &   1.077   &   -20.594   &   -19.935   &   2.188   &   8.663   &      0.027   &   A   \\ 
1-44210      &   8714-6101    &   07:54:51.899   &   45:49:21.20   &   0.745   &   4.439   &   -19.915   &   -19.432   &   1.969   &   7.304   &      0.022   &   C   \\
1-230177$^b$ &   8942-6101    &   08:19:35.486   &   26:21:45.59   &   1.841   &   1.950   &   -18.364   &   -17.371   &   3.845   &   6.060   &      0.020   &   A   \\
1-556820     &   10219-12702  &   08:22:01.425   &   21:20:34.35   &   1.191   &   8.519   &   -20.003   &   -19.113   &   3.145   &   8.791   &      0.022   &   C   \\ 
1-352635$^a$ &   10494-12705  &   08:24:31.862   &   54:51:14.00   &   1.687   &   19.67   &   -20.945   &   -20.168   &   2.895   &   14.888  &      0.025   &   C   \\
1-461292     &   8241-6101    &   08:28:11.646   &   17:22:28.77   &   2.056   &   0.500   &   -18.096   &   -17.747   &   1.545   &   5.818   &      0.021   &   C   \\
1-384848     &   9494-9102    &   08:29:44.361   &   22:25:27.83   &   1.502   &   11.55   &   -20.579   &   -19.930   &   2.178   &   7.167   &      0.025   &   C   \\
1-567155     &   8249-12704   &   09:09:30.604   &   45:57:08.46   &   2.087   &   29.96   &   -21.340   &   -20.465   &   2.889   &   12.161  &      0.027   &   A   \\
1-137908     &   8249-12703   &   09:18:14.205   &   45:39:06.08   &   1.473   &   30.47   &   -21.437   &   -20.630   &   2.770   &   16.087  &      0.027   &   A   \\
1-155926     &   8439-12702   &   09:26:09.434   &   49:18:36.72   &   5.325   &   27.63   &   -21.202   &   -20.145   &   4.695   &   8.688   &      0.027   &   A   \\
1-605884     &   8439-12703   &   09:28:32.574   &   50:47:37.04   &   0.645   &   14.66   &   -20.641   &   -19.689   &   3.107   &   11.824  &      0.025   &   C   \\
1-138102     &   8252-6102    &   09:38:13.909   &   48:23:17.89   &   1.376   &   12.73   &   -20.487   &   -19.717   &   2.570   &   6.603   &      0.026   &   A   \\
1-257412     &   8945-12705   &   11:37:16.260   &   45:45:26.64   &   1.133   &   6.067   &   -19.847   &   -19.222   &   1.752   &   11.586  &      0.025   &   A   \\
1-258315$^a$ &   8261-6102    &   12:10:49.285   &   44:30:45.37   &   2.002   &   9.398   &   -20.097   &   -19.257   &   2.705   &   5.550   &      0.023   &   A   \\
1-209199     &   8485-9102    &   15:36:26.766   &   47:51:17.23   &   2.260   &   16.08   &   -20.677   &   -19.767   &   2.961   &   7.931   &      0.026   &   B   \\ 
1-211103     &   8550-6103    &   16:30:33.285   &   39:49:50.62   &   3.676   &   14.46   &   -20.510   &   -19.616   &   2.996   &   6.467   &      0.025   &   B   \\
1-251686     &   8335-12705   &   14:27:02.366   &   39:57:25.96   &   0.733   &   16.52   &   -20.915   &   -20.177   &   2.413   &   13.532  &      0.025   &   B   \\
1-564490     &   8604-12704   &   16:28:39.485   &   40:07:25.37   &   1.030   &   14.10   &   -20.667   &   -19.901   &   2.739   &   10.620  &      0.026   &   B   \\
1-135030$^a$ &   8603-12704   &   16:31:34.530   &   40:33:56.15   &   2.595   &   31.90   &   -21.327   &   -20.343   &   3.728   &   9.510   &      0.027   &   B   \\
1-94066      &   8484-1902    &   16:34:46.011   &   45:19:27.52   &   0.941   &   10.65   &   -20.579   &   -19.836   &   2.826   &   12.537  &      0.025   &   B   \\
1-23979      &   7991-3702    &   17:12:38.101   &   57:19:20.72   &   2.967   &   8.923   &   -19.883   &   -18.724   &   4.832   &   4.332   &      0.027   &   B   \\
1-25632      &   8611-9101    &   17:22:41.792   &   59:51:06.72   &   3.310   &   14.10   &   -20.518   &   -19.587   &   3.349   &   7.198   &      0.027   &   B   \\
1-246549     &   8597-12703   &   15:01:33.354   &   49:06:44.75   &   2.332   &   18.66   &   -20.916   &   -19.957   &   3.870   &   10.677  &      0.026   &   B   \\
1-569169     &   8602-12701   &   16:28:11.561   &   39:49:18.84   &   5.960   &   47.52   &   -21.644   &   -20.618   &   4.722   &   17.457  &      0.027   &   B   \\
1-546807     &   7957-12704   &   17:09:42.583   &   36:24:53.82   &   2.241   &   5.955   &   -20.388   &   -19.832   &   2.230   &   12.495  &      0.028   &   B   \\
\hline
\multicolumn{12}{l}{$^a$VPH observations in Mg I region; $^b$ low-mass AGN host; $^c$ A = 14-15,18-21 January 2018, B = 8-9 June 2018, C = 30 November and 1-3 December 2018.}
\end{tabular}
\end{table*}

%%%%%%%%%%%%%%%%%%%%%%%%%%%%%%%%%%%%%%%%%%%%%%%%%%

% Don't change these lines
\bsp	% typesetting comment
\label{lastpage}
\end{document}